\documentclass[twocolumn]{aastex63}
\usepackage{xcolor}

\begin{document}

\title{Wave-driven mass loss of stripped envelope massive stars: progenitor-dependence, mass ejection, and supernovae}


\shorttitle{Wave Transport and Energy Deposition}
\shortauthors{Leung, Wu and Fuller}

\author[0000-0002-4972-3803]{Shing-Chi Leung}
\affiliation{TAPIR, Walter Burke Institute for Theoretical Physics, 
Mailcode 350-17, Caltech, Pasadena, CA 91125, USA}

\author[0000-0003-2872-5153]{Samantha Wu}

\affiliation{TAPIR, Walter Burke Institute for Theoretical Physics, 
Mailcode 350-17, Caltech, Pasadena, CA 91125, USA}

\author[0000-0002-4544-0750]{Jim Fuller}

\affiliation{TAPIR, Walter Burke Institute for Theoretical Physics, 
Mailcode 350-17, Caltech, Pasadena, CA 91125, USA}


\correspondingauthor{Shing-Chi Leung}
\email{scleung@caltech.edu}

\received{21 Septmber 2021}  
\accepted{1 October 2021}   
\submitjournal{Astrophysical Journal}

\date{\today}

\begin{abstract}

The discovery of rapidly rising and fading supernovae powered by circumstellar interaction has suggested the pre-supernova mass eruption phase as a critical phenomenon in massive star evolution. It is important to understand the mass and radial extent of the circumstellar medium (CSM) from theoretically predicted mass ejection mechanisms. In this work, we study the wave heating process in massive hydrogen-poor stars, running a suite of stellar models in order to predict the wave energy and pre-explosion time scale of surface energy deposition. We survey stellar models with main sequence progenitor masses from 20--70 $M_{\odot}$ and metallicity from 0.002 to 0.02. Most of these models predict that less than $\sim \! 10^{47} \, {\rm erg}$ is deposited in the envelope, with the majority of the energy deposited in the last week of stellar evolution. This translates to CSM masses less than $\sim \! 10^{-2} \, M_\odot$ that extend to less than $\sim \! 10^{14} \, {\rm cm}$, too small to greatly impact the light curves or spectra of the subsequent supernovae, except perhaps during the shock breakout phase. However, a few models predict somewhat higher wave energy fluxes, for which we perform hydrodynamical simulations of the mass ejection process. Radiative transfer simulations of the subsequent supernovae predict a bright but brief shock-cooling phase that could be detected in some type Ib/c supernovae if they are discovered within a couple days of explosion.

\end{abstract}



\section{Introduction}
\label{sec:intro}

\subsection{Rapid Transients}

Type Ib/c supernovae are caused by the explosion of a massive star in which the surface H-envelope has previously been lost through winds or mass transfer in a binary system \citep[see e.g.,][for a recent review]{Yoon2015}.
The launch of Zwicky Transient Factory (ZTF) has led to more detections of supernovae at hours to days after explosion. High-cadence monitoring also allows for detections of pre-supernova outbursts, e.g. in SN 2018gep at about two weeks before its final explosion \citep{Ho2020}. Similar outbursts have been observed in both Type II, Type Ib/c and Type Ibn supernovae such as SN 2007bg \citep{Milisavljevic2013}, SN 2008D \citep{Modjaz2009, Sivrski2014}, SN 2010mc \citep{Ofek2013}, PTF13efv \citep{Ofek2016}, SN 2015bh \citep{EliasRosa2015}, SN 2015U \citep{Shivvers2016}, and SN 2015G \citep{Shivvers2017}. SN 2009ip \citep{Mauerhan2013, Ofek2009} provides evidence of a star resembling a Luminous Blue Variable (LBV) going supernova, with a clear mass outburst 3 years before the real explosion. These outbursts have demonstrated the possibility that a massive star can lose mass dynamically, besides its stellar wind mass loss. 

Pre-supernova outbursts also connect to the formation of circumstellar medium (CSM) around the exploding star, which has been a challenge in stellar evolution theory. For very massive stars (Zero-Age Main-Sequence mass $M_{\rm ZAMS} = 80 - 140~M_{\odot}$), the electron-positron pair 
instability drives explosive O-burning and mass ejection, which accounts for an outburst of $\sim 0.1$ to tens of $M_\odot$ \citep{Woosley2019, Leung2019PPISN, Renzo2020}. However, for lower-mass stars, the mechanisms at play are not clear.

The observed transients with a rapid rise time (from a few days to $\sim10$ days) are usually associated with the existence of some shock interaction between the ejecta and the CSM. The interaction picture has been suggested to explain many bright and rapid transients, such as SN 2006gy \citep{Woosley2007, Blinnikov2010}, iPTF14hls \citep{Woosley2020}, AT2018cow \citep{Leung2020COW}, and SN 2018gep \citep{Leung2021SN2018gep}. 

\subsection{Dynamical Evolution of Massive Stars}


Late-phase nuclear burning of massive stars is rapid and strong, which can drive vigorous convective flow within the C- and O-burning regions. The convection triggers wave generation outside the convection zone, and some of these waves leak through the evanescent zones inside the star and propagate outward \citep{Quataert2012}. Escaped waves with sufficient energy can form shocks or dissipate via radiative diffusion when they approach the stellar surface, where they deposit their energy as an extra thermal energy source \citep{Shiode2014}. Even though the relative amount of energy which can successfully leak is small compared to the whole stellar energy budget, in some cases it may be sufficient to eject the outermost matter from the H- and He-envelope in a H-rich \citep{Fuller2017} or H-poor \citep{Fuller2018} star. The exact amount of mass ejection depends on the energy budget.

The energy injection can also change the near-surface structure of the star \citep{Owocki2019, Kuriyama2020,Leung2020RG} and hence its observable optical appearance \citep{Kuriyama2021}. Additionally, the circumstellar medium can greatly affect the light curve of the subsequent supernova \citep{Suzuki2019}.  The outburst mass and energy expected from wave heating are approximately capable of matching some well observed transients such as SN 2000kf \citep{Ouchi2021} and SN 2018gep \citep{Leung2021SN2018gep}.


%

\subsection{Motivation and Outline}

In \cite{Leung2021SN2018gep} we used a parameterized wave model to study how the stellar envelope of a H-poor star responds to the wave energy deposition. Depending on the energy deposition and duration, we estimated that the typical mass loss can reach $\sim 10^{-5}$ -- $10^{-2} ~M_{\odot}$. In this work, we examine the energy deposition computed from realistic stellar models that self-consistently compute the wave flux escaping from the stellar core.

In Section \ref{sec:methods}, we describe the numerical methods for computing the wave heating rate from stellar models, the following hydrodynamical evolution, and radiative transfer simulations. In Section \ref{sec:results}, we describe the wave heating rates from our suite of stellar models, and how it depends on stellar parameters such as mass and metallicity. Section \ref{sec:response} presents hydrodynamical simulations of the stellar response to wave heating, while Section \ref{sec:radtrans} shows light curve models of the subsequent supernova. In Section \ref{sec:discussion} we discuss implications for transients and comparisons with recent work in the literature, and we conclude in Section \ref{sec:conclusion}.

\section{Methods}
\label{sec:methods}

We use the stellar evolution code MESA (Modules for Experiments in Stellar Astrophysics) \citep{Paxton2011, Paxton2013, Paxton2015, Paxton2017, Paxton2019} version 8118. The stellar evolutionary model assumes the default mixing length index, exponential overshooting parameter of 0.025 and the \textit{Dutch} wind formulae \citep{Vink2000, Vink2001} with a coefficient of $\eta=0.5$. The necessary configuration files and extra subroutines are available on Zenodo\footnote{MESA run files are uploaded to Zenodo via the link \url{https://doi.org/10.5281/zenodo.5542375}.}.

Our supernova radiative transfer models uses the Supernova Explosion Code (SNEC) \citep{Morozova2015}. The code is based on the prototype reported in \cite{Bersten2011, Bersten2013}, which solves for the bolometric radiative transfer assuming blackbody radiation, with a realistic opacity table taking inputs of density, temperature and chemical composition.

In this work, each stellar evolutionary model is prepared by five steps. (1) We run the model until the core hydrogen is exhausted. (2) We remove the H-envelope by relaxing the stellar mass to the helium core mass, mimicking mass stripping via a companion star. (3) We continue our evolutionary model until core-collapse, now with the wave generation subroutine switched on to record the wave energy escaping from the core, but without depositing the energy in the envelope to prevent numerical difficulties. (4) For a limited set of models from (3), we add wave heat to the envelope and use the hydrodynamics module of MESA to simulate how the envelope expands and capture the mass ejection. (5) We use SNEC to compute the optical signal of the final explosion for the models from (4).

To calculate the wave heating rate of the envelope, we follow the formalism outlined in \cite{Fuller2018}, updated with the more realistic wave spectrum and calculation of nonlinear wave breaking described in \cite{Wu2020}. In each step we calculate how much energy generated in the convective core can successfully pass through all evanescent layers and reach the surface. We first locate the position and luminosity of individual actively burning core and shells. We classify the core and shells by the dominant element being burned, e.g. helium (He), carbon (C), oxygen (O), neon (Ne), and silicon (Si). We then integrate over the convective burning shell to estimate the 
wave energy flux and typical wave frequency generated by each layer.

Next, we extract the amount of escaped energy by calculating the neutrino damping attenuation factor $f_{\nu}$ and fraction of energy transmitted through the evanescent zones. The latter is approximately given by the probability of transmission through the thickest evanescent region, or equivalently the minimum transmission coefficient $T^2_{\rm min}$. For each angular wavenumber $l$, the fraction of energy which can escape is given by 
\begin{equation}
    f_{{\rm esc,}l} = \left( 1 + \frac{f_{\nu} - 1}{T^2_{\rm min,l}} \right)^{-1}
\end{equation}
The rate of energy escape to the envelope is then $L_{{\rm heat}} = \sum_{l=1}^{10} f_{{\rm esc},l} \dot{E}_l$, with $\dot{E}_l$ being the wave power put into each $l$ by convection. We assume the same angular wavenumber spectrum as \cite{Wu2020}.

We also crudely account for wave attenuation via non-linear wave breaking in the core, as discussed by \cite{Wu2020}, which is especially important in stripped-envelope stars.
For a given wave luminosity $L_{\rm wave}$ which can be transmitted through the evanescent region, we calculate the nonlinear coefficient $|k_r \xi_r|_l$ by finding
\begin{equation}
    |k_r \xi_r|_l = \left[ \frac{2}{T^2_{{\rm min},l}} \frac{L_{{\rm heat},l} N [l (l+1)]^{3/2}}{4 \pi \rho r^5 \omega^4} \right]^{1/2},
\label{eq:nonlinearity}
\end{equation}
where $T^2_{{\rm min},l}$ is the transmission coefficient described above and
$L_{{\rm heat},l}$ is the escaped energy flux (luminosity) for a given wavenumber $l$. Quantities $N$, $\rho$ and $r$ and $\omega$ are the local Brunt-V{\"a}is{\"a}l{\"a} frequency, density, radius, and wave frequency, respectively. The effective energy escape rate for each $l$ is given by 
\begin{equation}
    L_{\rm esc,~eff,l} = L_{{\rm esc},l} / |k_r \xi_r|_l^2.
\end{equation}
We refer the interested readers to the said references for the full description.

\section{Results}
\label{sec:results}

\subsection{Evolution of an Example H-poor Model}

We first examine how the wave energy transport occurs in a typical H-poor model. To illustrate this, we take an example of $M_{\rm ZAMS} = 60~M_{\odot}$, with $Z = 0.02$. This model evolves to form a 27.2 $M_{\odot}$ He-core, and later wind-driven mass loss during core helium burning leads to a final mass of $M_{\rm pre-SN} = 17.5~M_{\odot}$. 

\begin{figure*}
\centering
\includegraphics*[width=8cm]{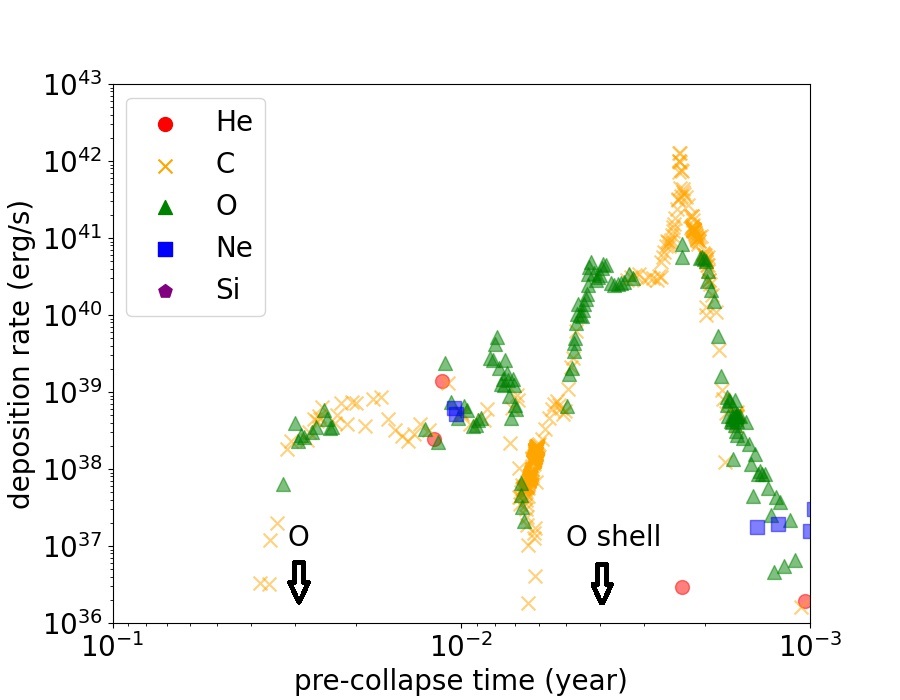}
\includegraphics*[width=8cm]{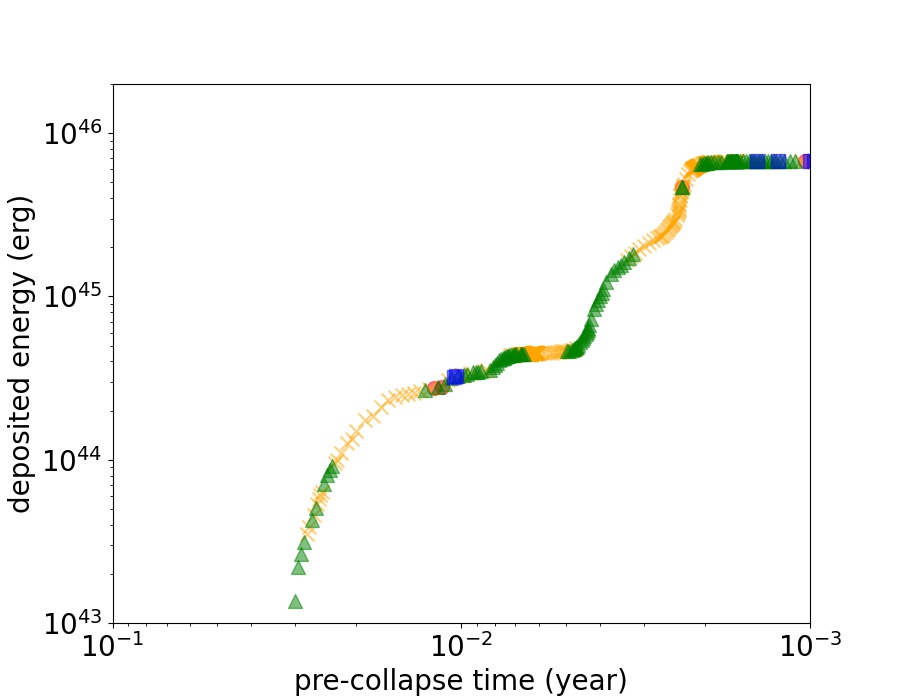}
\includegraphics*[width=8cm]{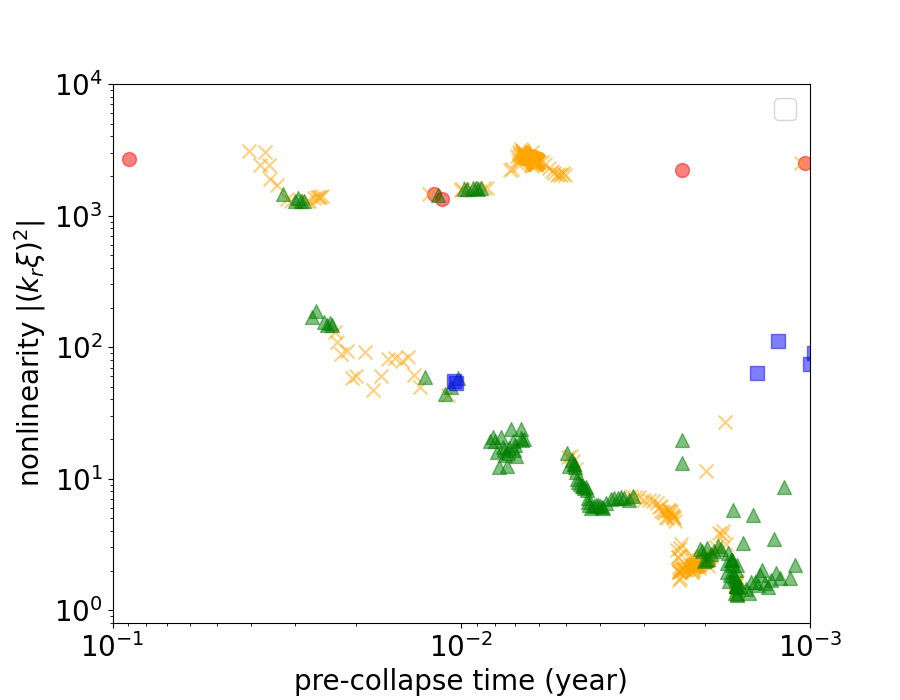}
\includegraphics*[width=8cm]{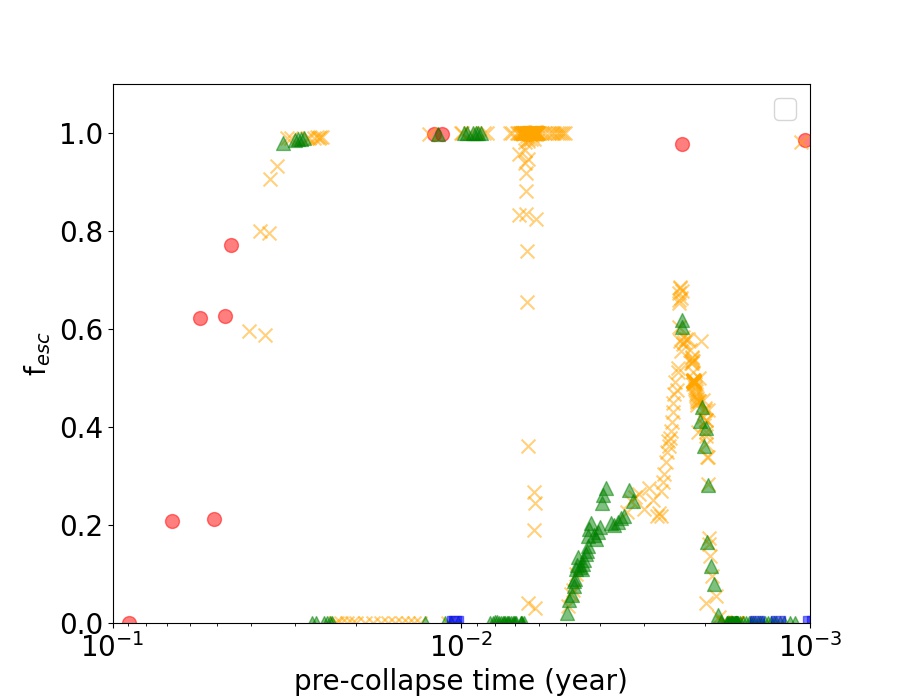}
\caption{
(top left panel) The total energy deposition rate in the envelope as a function of time before core-collapse for an H-poor model with $M_{\rm ZAMS} = 60~M_{\odot}$ and $Z = 0.02$. The colour corresponds to the type of convective burning region which contributes the most energy, as indicated in the legend. Arrows with ``O" and ``O-shell" labels mark the onset of O-core and O-shell burning. (top right panel) The cumulative deposited energy against time. (bottom left panel) The non-linearity attenuation factor for $l=1$ waves in the core. (bottom right panel) The escape fraction of $l=1$ waves.}
\label{fig:M60Z002D05noH_evol}
\end{figure*}

In Figure \ref{fig:M60Z002D05noH_evol} we plot the wave energy transport history of this model. We plot the wave energy deposition rate in the envelope (top left panel), cumulative deposited energy (top right panel), nonlinearity (bottom left panel), and the energy escape fraction (bottom right panel). In all the four panels, we mark the data points by colour to specify which burning zone contributes the most amount of energy, with He, C, O, Ne, and Si represented by red, orange, green, blue and purple respectively.

From the energy deposition rate, we observe that wave heating first becomes significant during core O-burning, about 0.05 years pre-explosion. C-shell burning and O-shell burning provide additional peaks in the escaped energy at $\sim 2-4 \times 10^{-3}$ year before collapse. After that, the energy escape rate sharply decreases. Si burning is almost insignificant to the energy budget up to $10^{-3}$ year before collapse. All told, roughly $10^{46} \, {\rm erg}$ of wave energy escapes to the envelope before core-collapse.


The escape fraction measures how much energy can reach the envelope after energy losses in the core and tunneling through evanescent regions. In this particular model, the escape fraction during core O-burning ($\sim \! 2 \times 10^{-2} \, {\rm yr}$ before core-collapse) is much lower than that during subsequent shell C- and O-burning. Hence, most of the energy from core O-burning is damped out via neutrino damping before being able to tunnel into the envelope.
The energy escape rate is also greatly decreased by the nonlinearity attenuation factor (equation \ref{eq:nonlinearity} squared). In this model, the core O-burning phase experiences a nonlinear attenuation of about 100, while that during C-shell and O-shell burning is about an order of magnitude lower. Hence, the wave heating rate jumps from $\sim 10^{39}$ erg during core O-burning to $\sim 10^{41}$ erg during C- and O-shell burning, demonstrating the importance of nonlinear suppression of wave energy transport.

\subsection{Comparison with H-rich model}

To understand the features in the H-poor model, it is necessary to discuss the wave transport feature in a H-rich model. To do so, we consider a H-rich model with $M_{\rm ZAMS} = 36~M_{\odot}$, which evolves to have a pre-explosion He core mass of 17.0 $M_{\odot}$.
This model has a similar pre-explosion He core mass as the H-poor model in the previous section, so its core evolves similarly during late stage burning.
This model transmits about $2 \times 10^{47}$ erg of wave energy to the surface, about an order of magnitude higher than the corresponding H-poor model with a similar He core mass. 

\begin{figure*}
\centering
\includegraphics*[width=8cm]{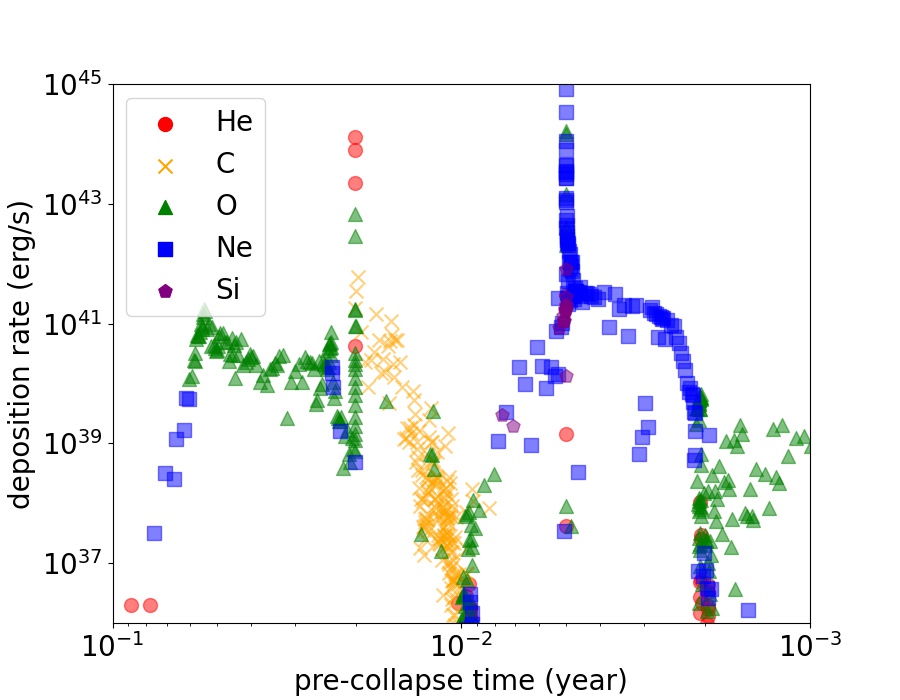}
\includegraphics*[width=8cm]{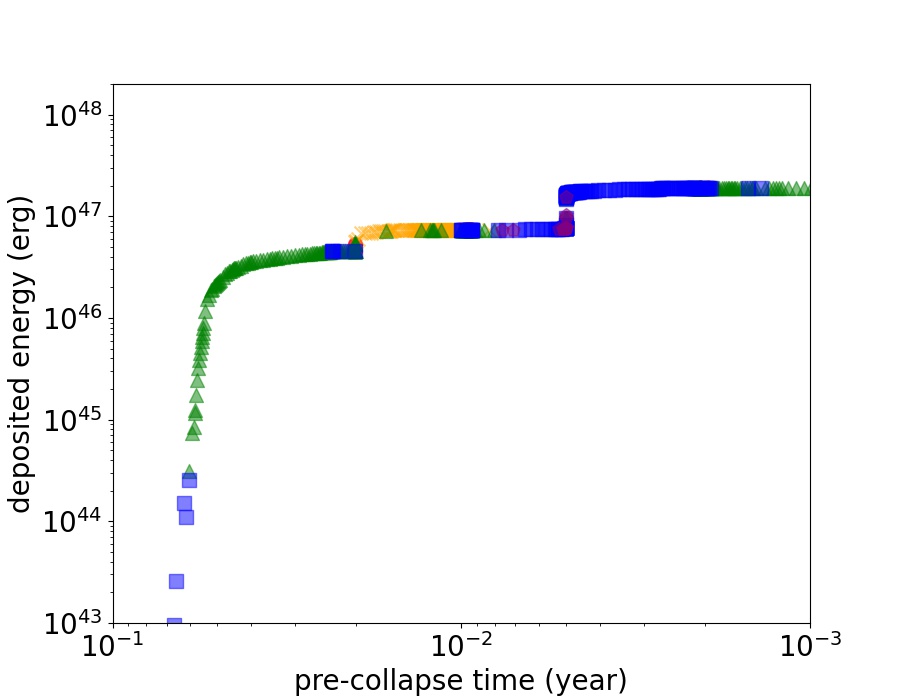}
\includegraphics*[width=8cm]{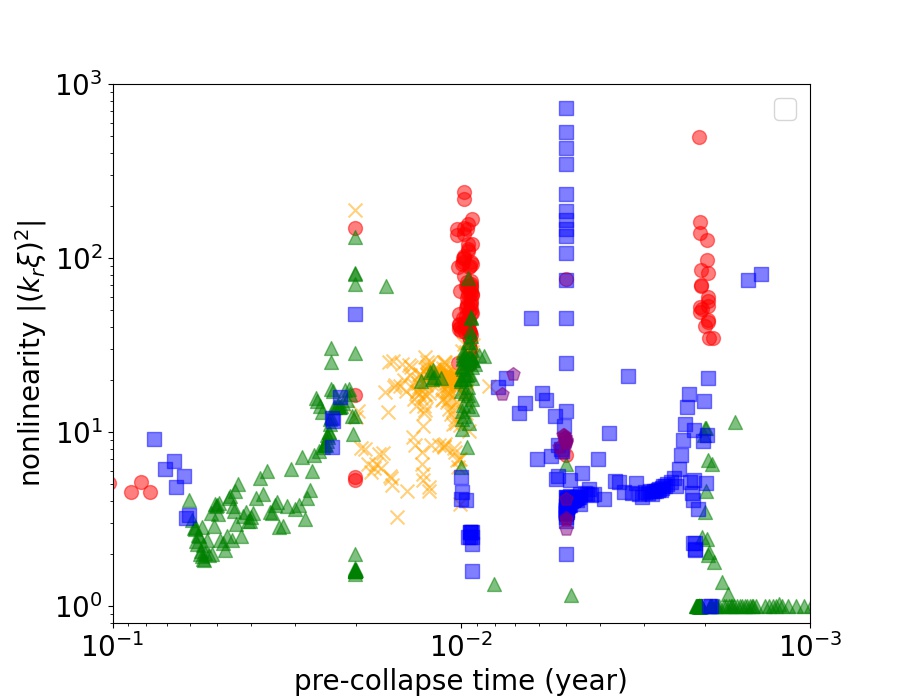}
\includegraphics*[width=8cm]{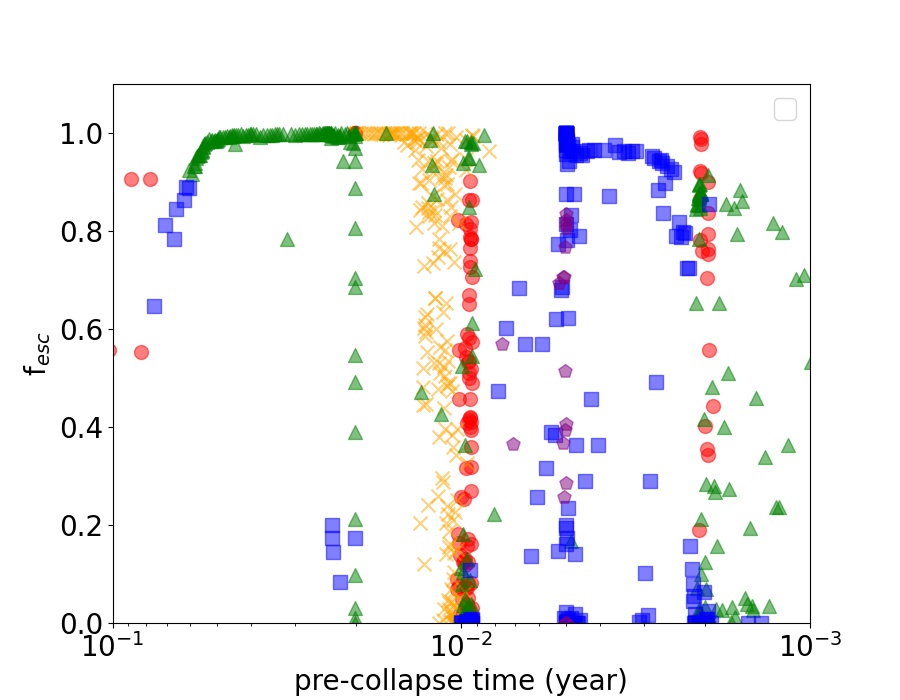}
\caption{
Same as Figure 1 except for an H-rich model with $M_{\rm ZAMS} = 36~M_{\odot}$ and $Z = 0.02$. 
}
\label{fig:M36Z002D05wH_evol}
\end{figure*}

The energy escape rate is much higher in the H-rich model during O-burning with a value $\sim \! 10^{41}$ erg s$^{-1}$, which is about two orders of magnitude higher than that of the H-poor model. The later C-shell and Ne-shell burning deposit energy with the same order of magnitude (See Figure \ref{fig:M36Z002D05wH_evol}). 
The primary reason for the larger heating is that the H-rich star has a larger escape fraction and much lower wave nonlinearity compared to the H-poor model with the same $M_{\rm pre-SN}$. 
In H-rich stars, the convective C-burning shell is usually narrower, such that the g mode cavity overlying the O-burning core is wider, and the evanescent region separating it from the envelope is smaller. 
Hence, a larger fraction of the waves escape the core, and a smaller fraction of the wave energy is lost to photon, neutrino, or non-linear damping. 

\begin{figure*}
\centering
\includegraphics*[width=18cm]{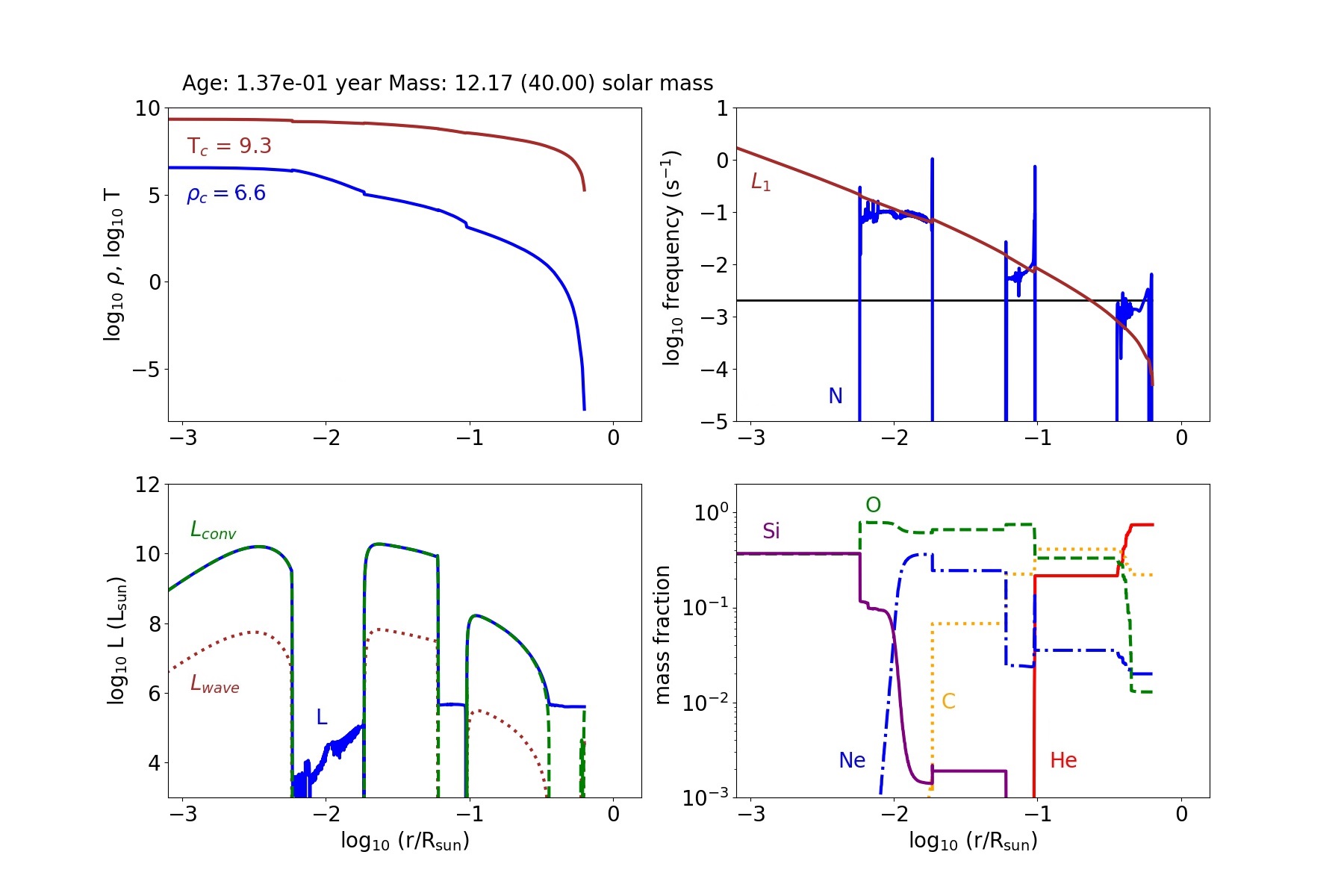}
\caption{
(top left panel) The density and temperature profiles for the H-poor model 
with $M_{\rm ZAMS} = 40~M_{\odot}$, $Z=0.02$, taken during core O-burning. (top right panel) A wave propagation diagram showing the Brunt-V\"ais\"al\"a frequency (blue), $l=1$ Lamb frequency (brown), and characteristic wave frequency (black). (bottom left panel) The total, wave and convective luminosities of the same model. (bottom right panel) The chemical abundance profile of the same model.}
\label{fig:M60Z002D05noH_profiles}
\end{figure*}

\begin{figure*}
\centering
\includegraphics*[width=18cm]{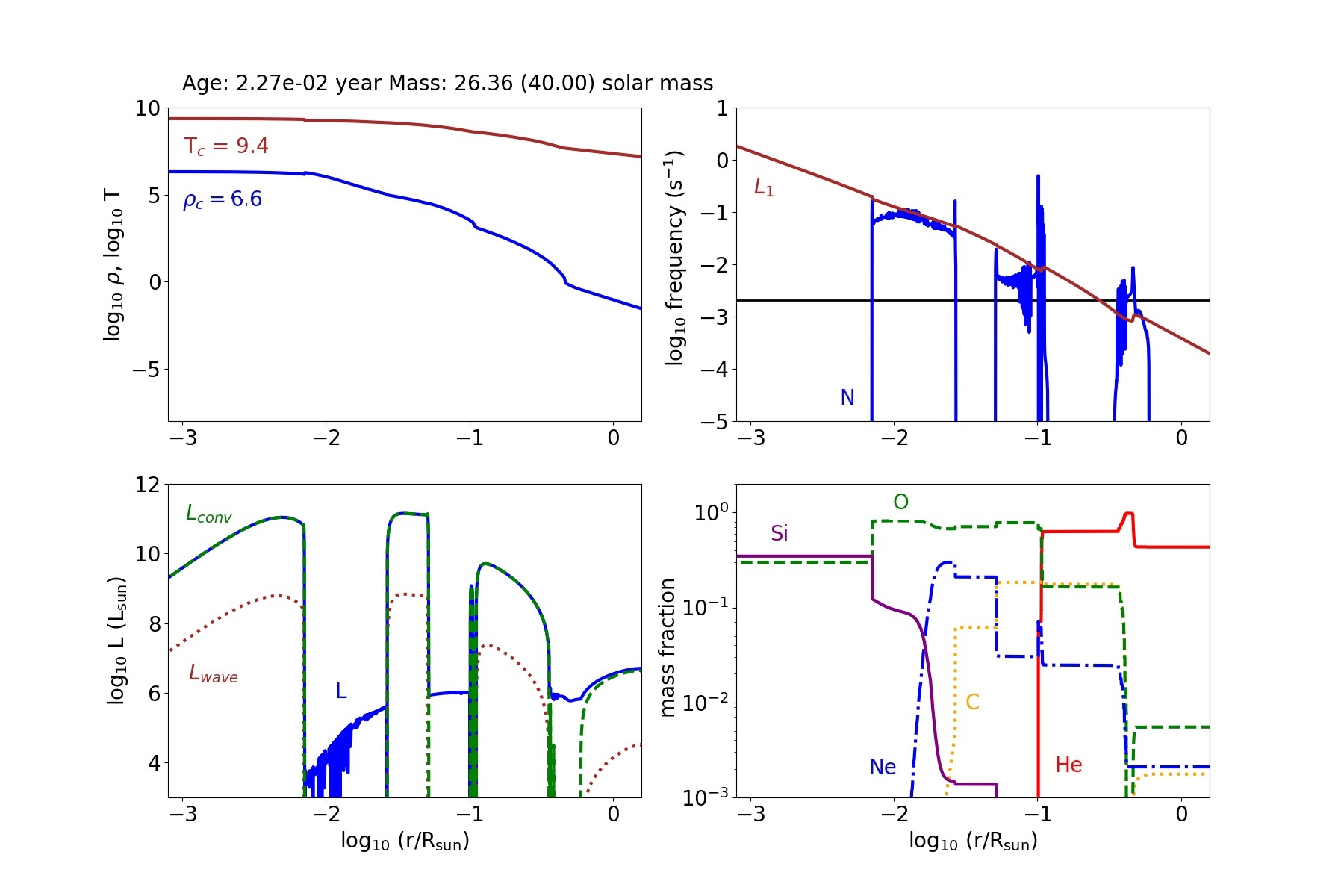}
\caption{
Same as Figure \ref{fig:M60Z002D05noH_profiles} but for the H-rich model with similar helium core mass.
}
\label{fig:M36Z002D05wH_profiles}
\end{figure*}

To illustrate this idea, we examine in Figures \ref{fig:M60Z002D05noH_profiles} and \ref{fig:M36Z002D05wH_profiles} the wave propagation diagrams for two models when core O-burning takes place. The models are chosen to have $M_{\rm ZAMS} = 40~M_{\odot}$ and $Z = 0.02$.
Most of the qualitative features in both H-rich and H-poor models are similar, in terms of the density and temperature profiles, and the luminosity profile. The main difference is the extensive H-envelope which appears in the H-rich model which extends to $\sim \! 1000~R_{\odot}$. The convective energy transport peaks at $\sim \! 10^{-3} - 10^{-1}~R_{\odot}$, and in both models the O-shell convective frequency is comparable, about $2 \times 10^{-3}$ s$^{-1}$.
The major difference comes the Brunt-V{\"a}is{\"a}l{\"a} frequency profiles. In the H-poor model, the convective C-burning shell is thicker, trapping more of the waves generated by the O-burning core beneath it. In the H-rich model, this convective shell is thinner, allowing the waves to more easily tunnel into the overlying radiative region between the C-burning and He-burning shells. Hence, in the H-rich models, more of the wave energy escapes from the core before it is damped.



\subsection{Distribution of Stellar Mass Loss}

\begin{figure*}
\centering
\includegraphics*[width=18cm, height=10cm]{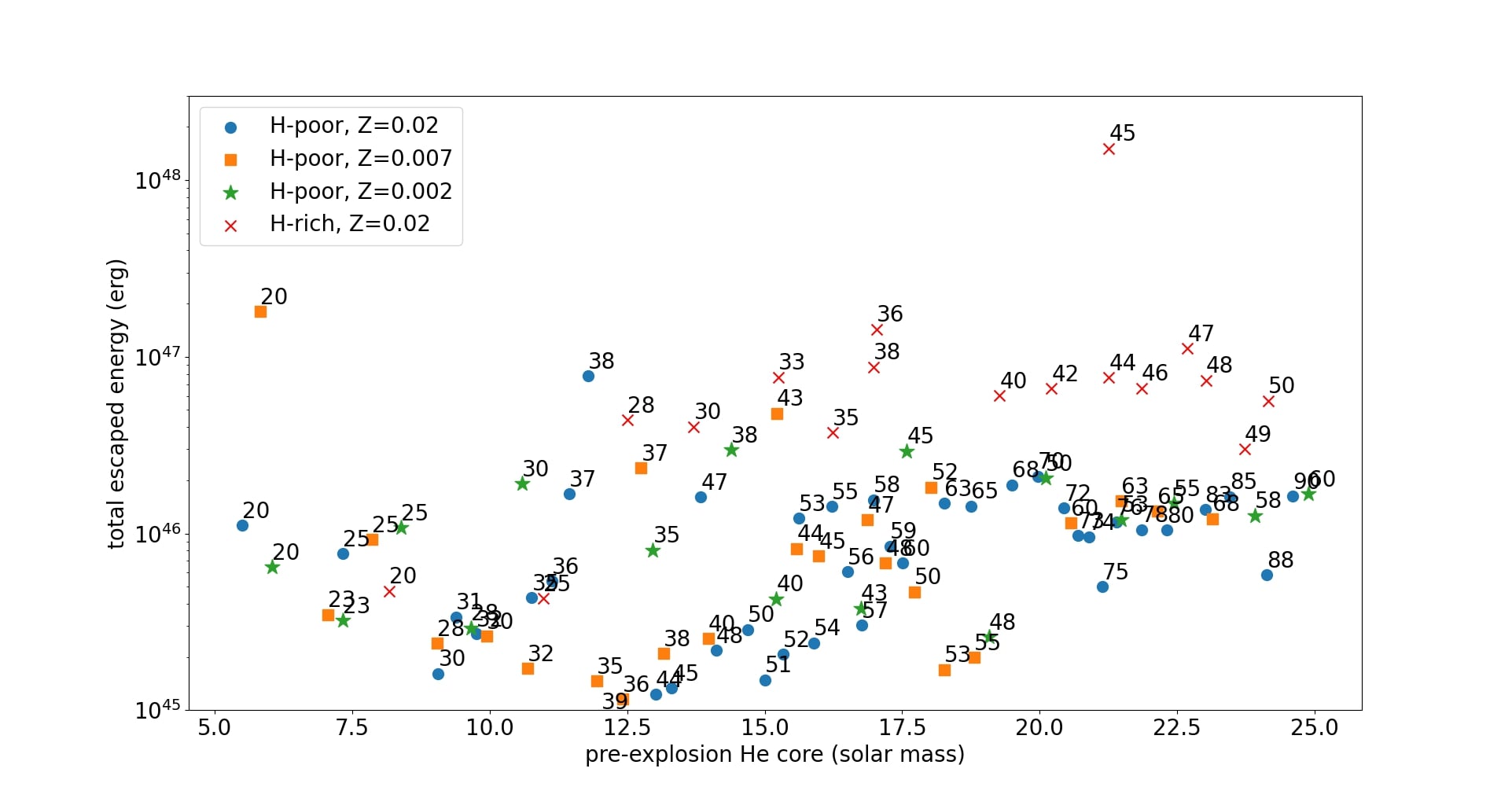}
\caption{
The time-integrated escaped energy from the core against pre-explosion mass for stellar models with mass and metallicity as parameters. H-rich models are included for comparison. The numbers correspond to the initial ZAMS mass.}
\label{fig:global_Edep}
\end{figure*}

We carry out an extensive survey of stellar models by varying two model parameters, initial ZAMS mass and metallicity. We choose a wide range of mass and metallicity to explore the potential mass loss driven by wave energy deposition. For lower metallicity models, we also lower the maximum mass simulated so that other major mass loss mechanisms, such as the mass loss driven by the pulsational pair-instability, are avoided. 

In Figure \ref{fig:global_Edep} we plot the time-integrated escaped energy of each stellar model against its pre-explosion He-core mass $M_{\rm pre-SN}$. We choose $M_{\rm pre-SN}$ as it is more closely related to the final evolution than $M_{\rm ZAMS}$. The total escaped energy is integrated from after He-core burning, up to $\sim 10^{-3}$ year before stellar collapse.
We have included models with a metallicity from 0.002 to 0.02. A subset of H-rich models are also included for comparison.

The clustering of the data points demonstrates multiple features of stellar evolution at different phases. The H-rich models have the highest escaped energy $E_{\rm esc}$. The energy range is consistent with previous work \citep{Wu2020} showing that H-rich stars often release about $10^{47}$ erg before explosion for a wide range of masses.
An outlier is the $M = 45~M_{\odot}$ model. This model has an extraordinarily high $E_{\rm esc} \sim \! 10^{48} \, {\rm erg}$. The origin of the high escaped energy is the convective shell merger phenomenon, as reported in previous work \citep{Wu2020}. During the C-shell burning, the growing C-burning shell merges with the overlying He-burning shell, dredging He-rich matter down into the actively burning C-shell. The influx of extra He can rapidly increase the nuclear reaction luminosity due to $\alpha-$capture reactions. As a result, the burning drives vigorous convection and wave generation.

On the other hand, H-poor models in general have a lower  $E_{\rm esc}$ by an order of magnitude ($\sim \! 10^{46}$ erg). Regardless of their initial metallicity, most of the data points cluster in a band that starts from about $10^{46}$ erg at $M_{\rm exp} \sim 5~M_{\odot}$ and then gradually decreases down to $10^{45}$ erg at $M_{\rm exp} \sim 12.5~M_{\odot}$. The band gradually increases and reaches an asymptotic energy $\sim \! 10^{46}$ erg again at $M_{\rm pre-SN} \sim 17.5 ~M_{\odot}$. 

A few outlying H-poor models have larger total escaped energies, such as the $M_{\rm ZAMS} = 38 \, M_\odot$ model which has $E_{\rm esc} \sim \! 10^{47} \, {\rm erg}$. 
The differences come from a number of physical processes which will be discussed in later sections. 

    

\begin{table}[]
    \centering
    \caption{Model outliers  with large wave energy deposition from Figure \ref{fig:global_Edep}. The right column lists the reason for the large wave energy, as well as the burning zone type that provides the most wave heat.}
    \begin{tabular}{c c c c c}
        \hline
        Group & $Z$ & $M_{\rm ZAMS} \, (M_\odot)$ & Stripped? & Origin \\ \hline
        A & 0.007 & 20 & Yes & low $k_r \xi_r$ \\
        & & & & high $f_{\rm esc}$ (Ne) \\ \hline 
        B & 0.02  & 37, 38, 47 & Yes & low $k_r \xi_r$ \\
          & 0.007 & 37, 43 & Yes & high $f_{\rm esc}$ (C/O) \\
          & 0.002 & 30, 38 & Yes & \\ \hline
        C & 0.02  & 45 & No & Shell merger \\
        & & & & (He/C) \\ \hline

    \end{tabular}
    
    \label{table:outliers}
\end{table}

\subsection{Distribution of deposition timescale}

\begin{figure*}
\centering
\includegraphics*[width=18cm, height=10cm]{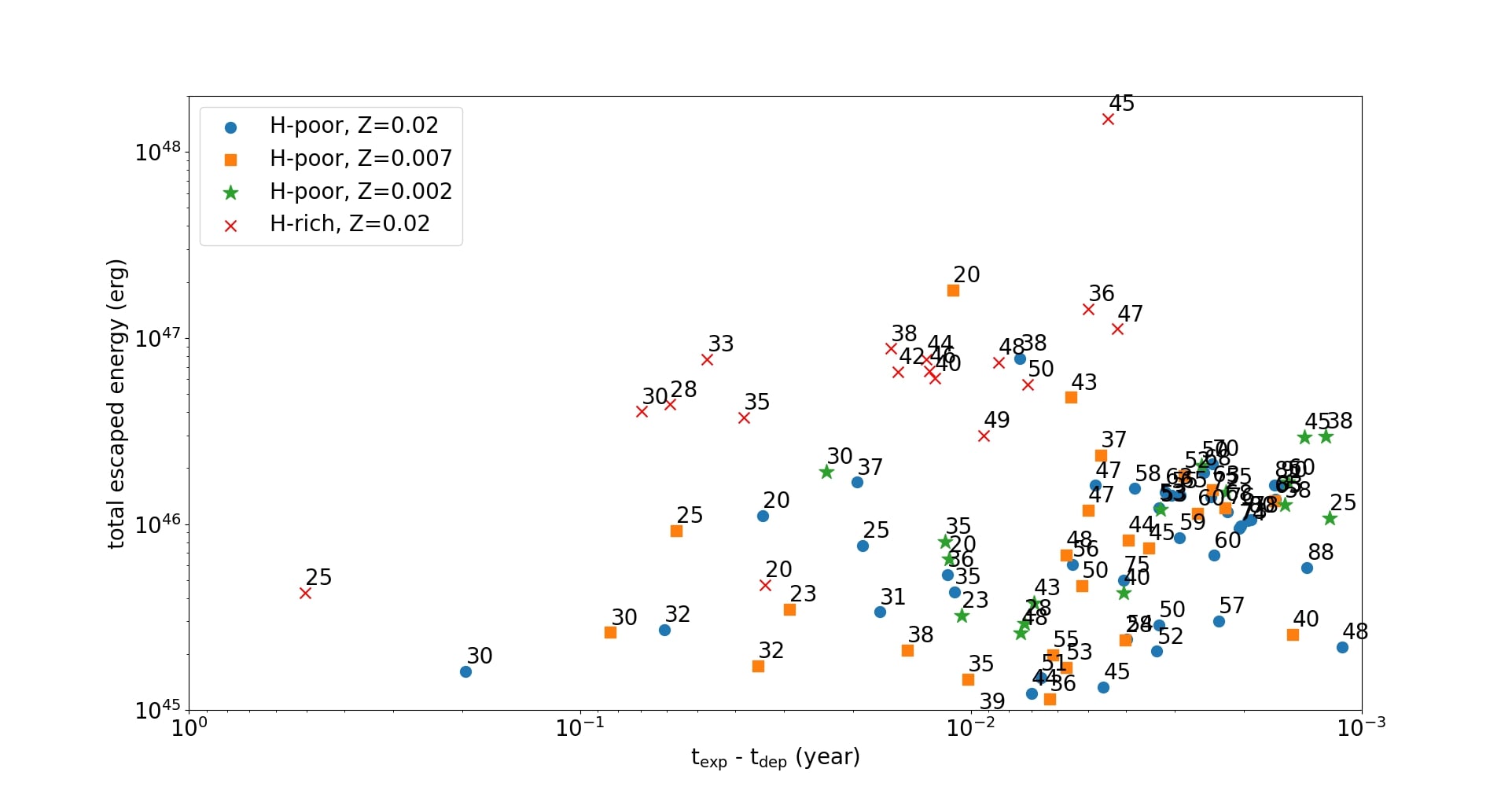}
\caption{
The time-integrated escaped energy from the convective core against the characteristic energy deposition time before core-collapse for stellar models in Figure \ref{fig:global_Edep}. H-rich models are included for comparison. The numbers correspond to the initial ZAMS mass.}
\label{fig:global_Edep_time}
\end{figure*}

In Figure \ref{fig:global_Edep_time} we plot the total escaped energy against deposition time $t_{\rm dep}$ for each model. We define the deposition time by the time before core-collapse at which 50\% of the total wave energy has escaped from the core. 
This corresponds approximately to the time before explosion of any pre-supernova mass ejection that occurs. However, whether or not the star ejects mass must be determined by hydrodynamic simulations of the envelope's response to wave heating (e.g., Section \ref{sec:response} and \citealt{Leung2021SN2018gep}).

Regardless of the star being H-rich or H-poor, the majority of models have deposition time scales from $10^{-3}$ to $10^{-1}$ years. Stars with higher pre-explosion helium core masses tend to have shorter deposition time scales, but there is large scatter and the value of $t_{\rm dep}$ is not correlated with the escaped energy. For example, models with $M_{\rm ZAMS} = 25 ~M_{\odot}$ have an outburst time as early as 0.09 year before explosion for the model with $Z = 0.007$, but it is as low as 0.01 year for $Z = 0.02$ and 0.001 year for $Z = 0.002$. Each represents a distinctive energy deposition history, and hence we expect there corresponding circumstellar environment can be very different. 

\begin{figure*}
\centering
\includegraphics*[width=8cm]{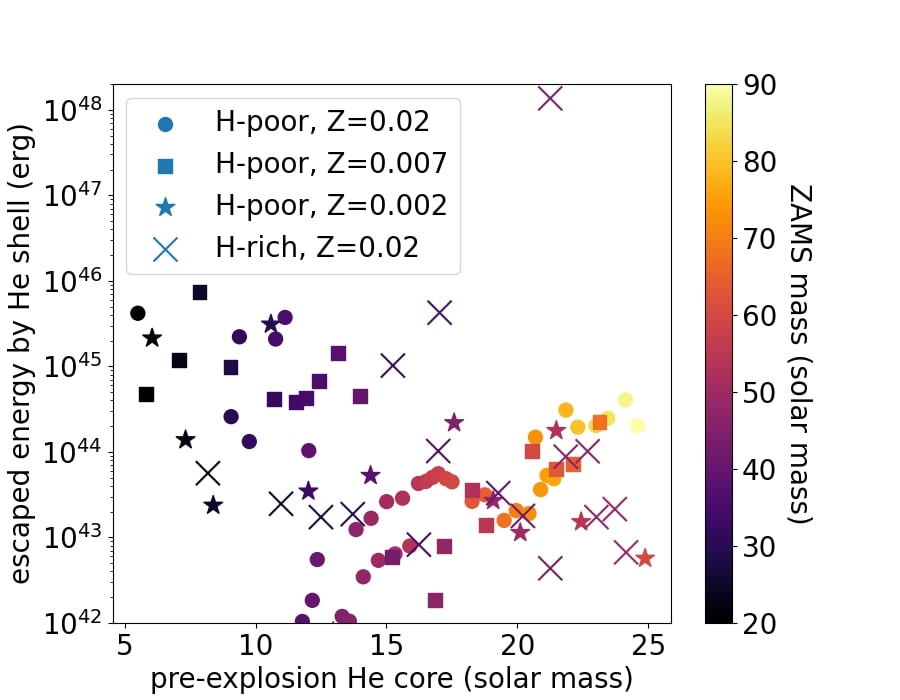}
\includegraphics*[width=8cm]{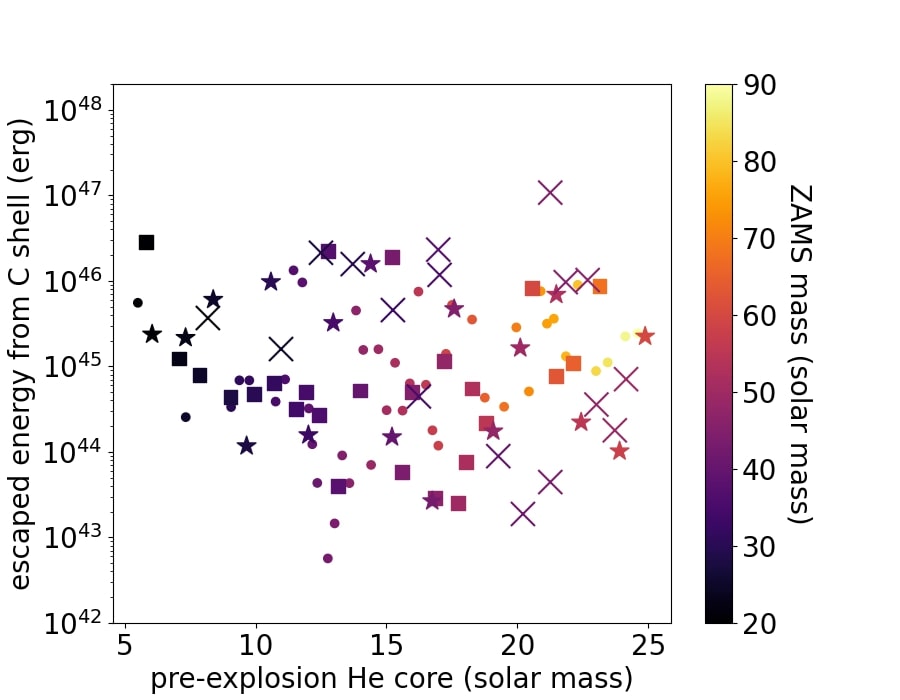}
\includegraphics*[width=8cm]{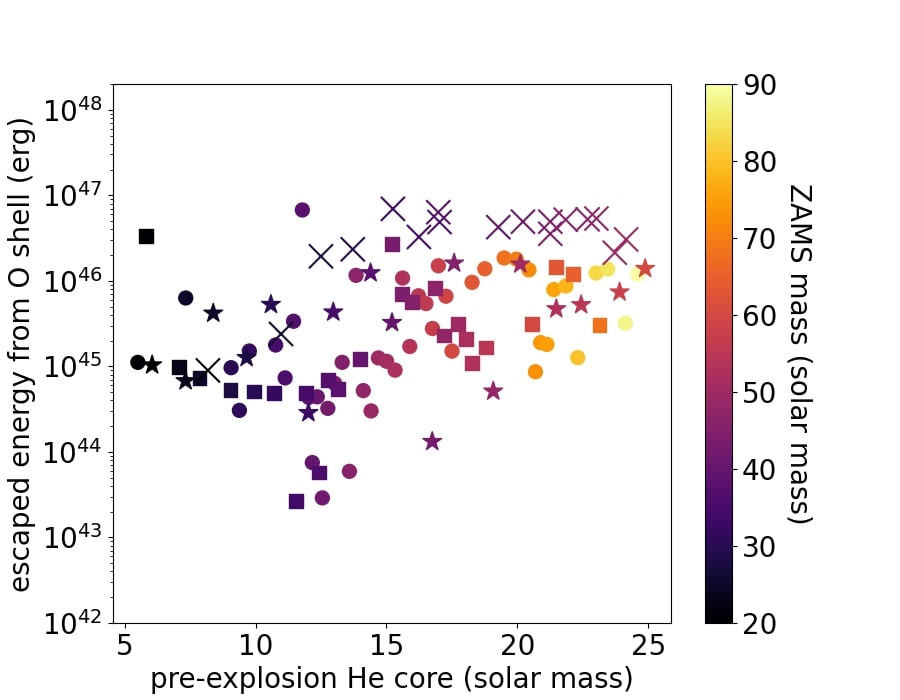}
\includegraphics*[width=8cm]{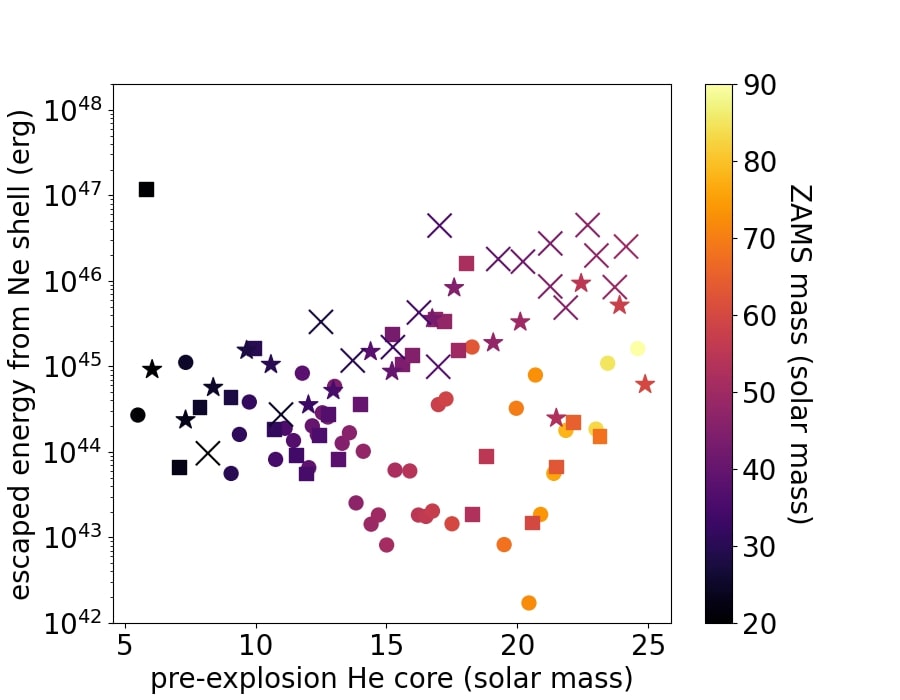}
\caption{
(top left panel) The wave energy transmitted to the envelope that is generated by the He-burning shell against $M_{\rm pre-SN}$ for both H-rich and H-poor models. (top right panel) Same as top left panel but for the C-burning shell. (bottom left panel) Same as top left panel but for the O-burning shell. (bottom right panel) Same as top left panel but for the Ne-burning shell.}
\label{fig:M25_Ldep_channel}
\end{figure*}

We also examine how each of the burning channels contributes to the total deposited energy in each model. In Figure \ref{fig:M25_Ldep_channel} we plot the energy deposited by waves generated from convective burning of He (top left panel), C (top right panel), O (bottom left panel) and Ne (bottom right panel). The Si channel in general contributes an insignificant amount of energy except just before core-collapse. 

During He burning, the majority of models scatters between $10^{43}$ -- $10^{45}$ erg, which is at most about $\sim 10\%$ of the deposited energy. 
The extremely high value for one H-rich model corresponds to the shell merger model described in the previous section.
For the C burning, the total escaped energy ranges from $10^{44}$ -- $10^{46}$ erg. There is no significant trend between the pre-explosion mass and the total escaped energy.

O burning often provides the most energy and has a narrower range between $\sim 10^{45}$ -- $10^{47}$ erg, but other types of burning are most important in a significant fraction of models.
We observe a narrow band for the H-rich models, which is consistent with previous work \citep{Wu2020} that O-burning provides the majority of wave energy to the envelope. Ne burning has a similar scatter as the C-shell with a range from $10^{43}$ --  $10^{46}$. H-rich models have significantly higher energy from Ne burning than H-poor models.

\subsection{A Case Study of a $M_{\rm ZAMS} = 25~M_{\odot}$ Star}

\begin{figure*}
\centering
\includegraphics*[width=5.5cm]{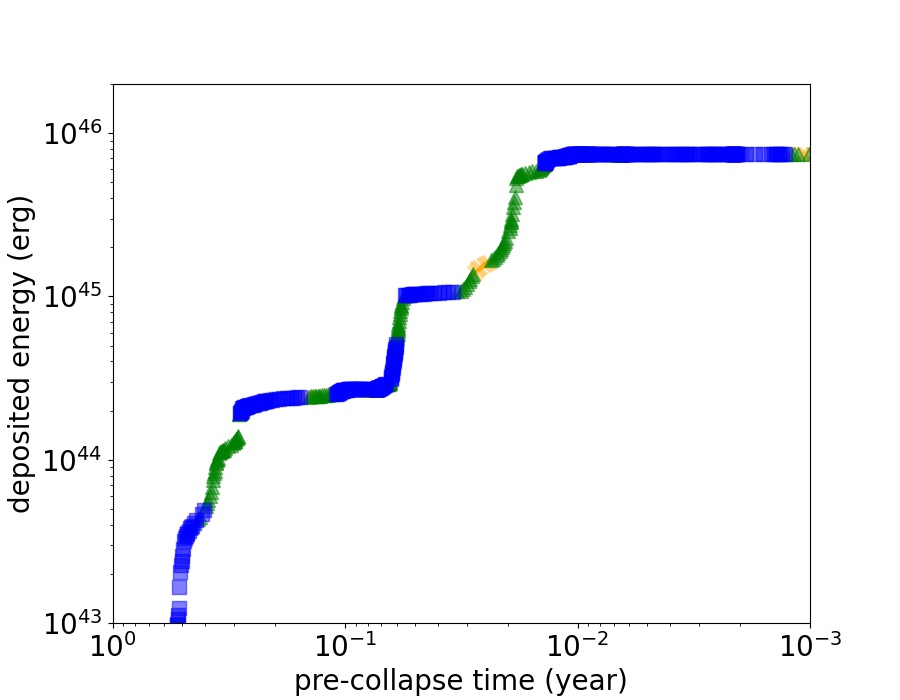}
\includegraphics*[width=5.5cm]{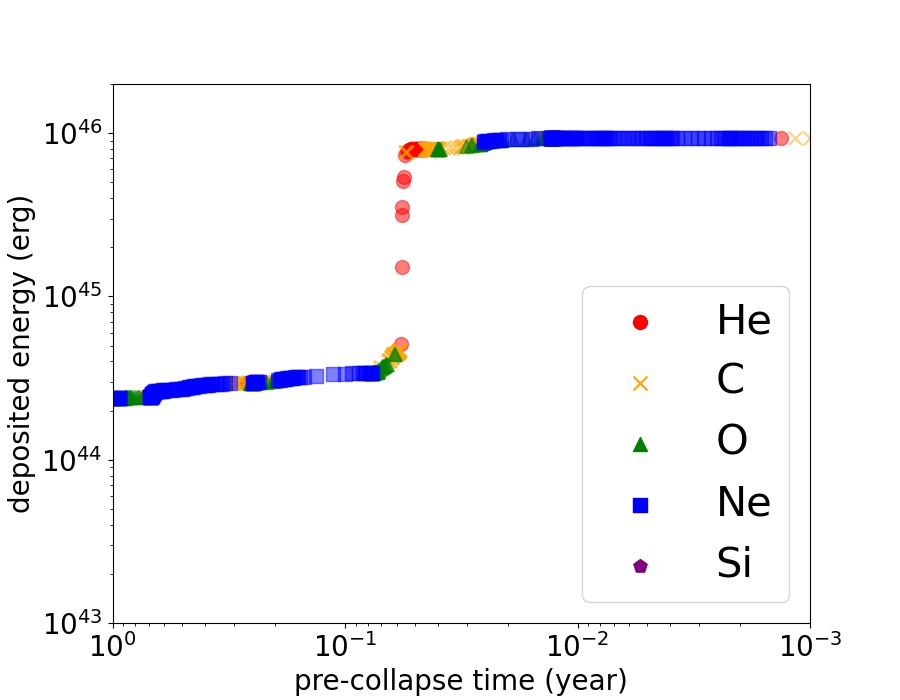}
\includegraphics*[width=5.5cm]{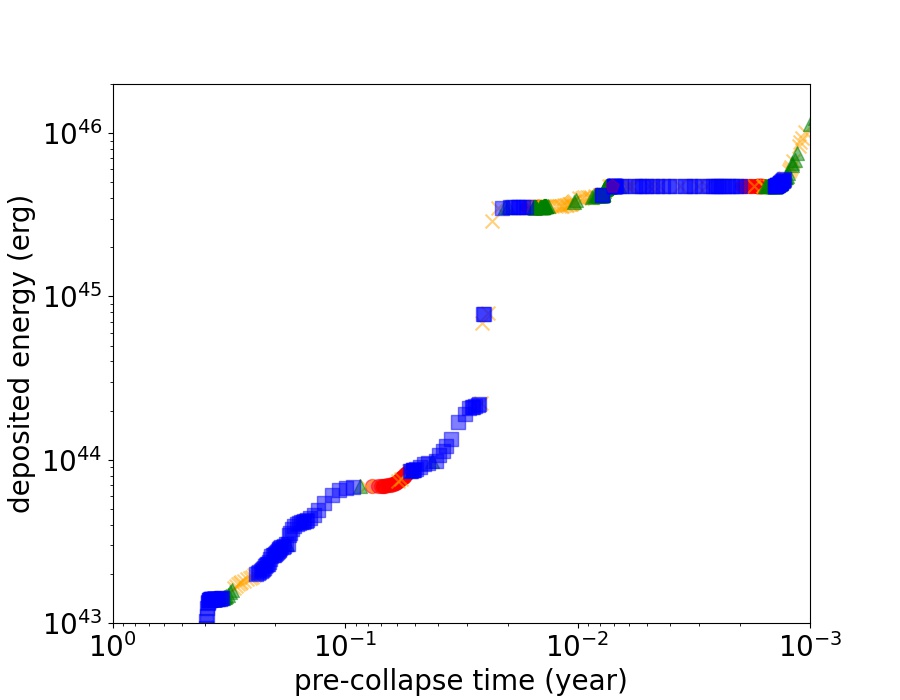}
\caption{
The cumulative deposited wave energy versus the pre-explosion time for the model with $M_{\rm ZAMS} = 25~M_{\odot}$ and $Z = 0.02$ (left panel), $Z = 0.007$, (middle panel), and $Z = 0.002$ (right panel).
}
\label{fig:M25_Edep_hist}
\end{figure*}

We next focus on a specific group of models with $M_{\rm ZAMS} = 25~M_{\odot}$ and $Z = 0.002$, 0.07 and 0.02. These models have a comparable total deposited energy, but each has a distinctive deposition duration. In Figure \ref{fig:M25_Edep_hist} we plot the cumulative energy as a function of pre-collapse time for the three models. It is evident that the timing of the wave heating episodes varies significantly between the models, even though they each end up with $\sim \! 10^{46}$ erg of wave heat. Moreover, the burning phase (e.g., helium, carbon, or oxygen burning) responsible for the majority of the heat is different in each model. Clearly, models with similar He core masses can differ greatly in both net wave energy and the deposition time scale. 



\subsection{Comparison of High and Low Energy Models}

\begin{figure*}
\centering
\includegraphics*[width=8cm]{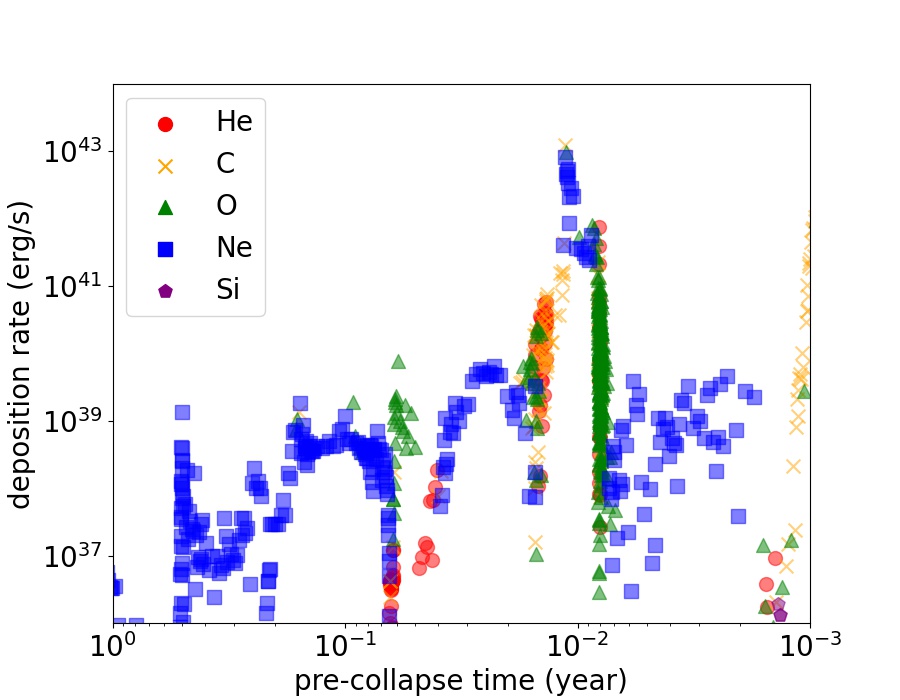}
\includegraphics*[width=8cm]{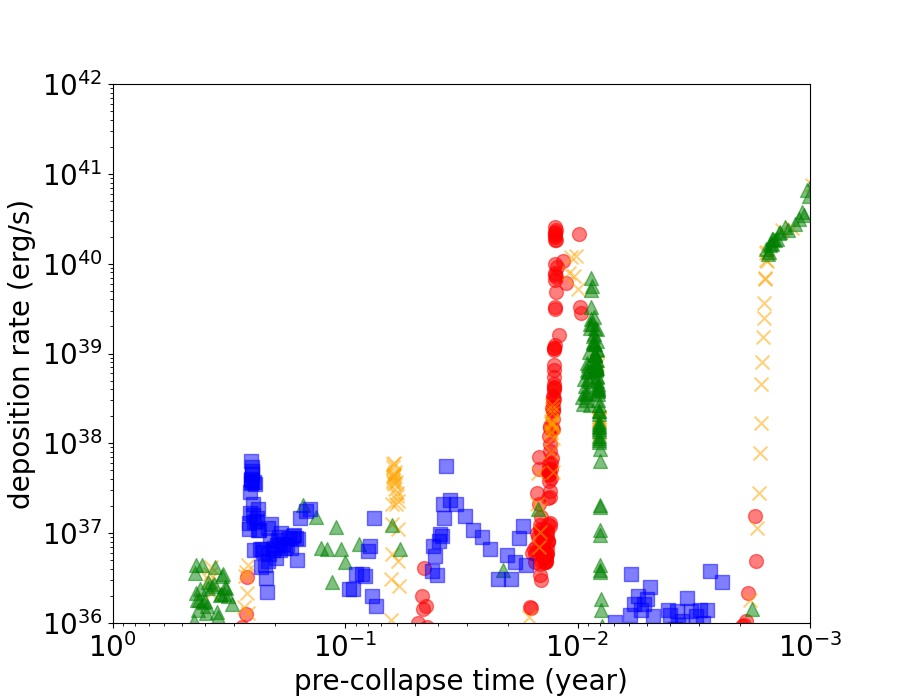}
\caption{
The wave energy escape rate against time for the high wave energy H-poor model with $M_{\rm ZAMS} = 20~M_{\odot}$ and $Z = 0.007$ (left panel) and the low wave energy H-poor model with $M_{\rm ZAMS} = 35~M_{\odot}$ and $Z = 0.007$ (right panel).
}
\label{fig:M20Z0007D05noH_evol}
\end{figure*}

In Figure \ref{fig:global_Edep} we show that there are outliers with more escaped wave energy than most models. Here we study them in detail. We compare two distinctive models: an H-poor model with $M_{\rm ZAMS} = 20~M_{\odot}$ and $Z = 0.007$ (high energy model), and an H-poor model with $M_{\rm ZAMS} = 35~M_{\odot}$ and $Z = 0.007$ (low energy model).
In Figure \ref{fig:M20Z0007D05noH_evol} we plot the wave heating rate of each model. The two models demonstrate qualitative similarities, but the wave heating rate is typically 2 orders of magnitude larger in the high-energy model. 

How most of the energy is released at $\sim \! 0.01$ year before collapse is different in the two models. Even though there is a sudden burst of wave energy from O- (He-) shell burning in the high (low) energy model, the corresponding escape fraction and nonlinearity during that period is very different. The high energy model maintains a large escape fraction and a low non-linearity ($\sim 1$) for about 0.003 year, while the low energy model has a small escape fraction and large non-linearity ($\sim 8$). This drastically impacts the energy which can successfully reach the envelope.


\begin{figure*}
\centering
\includegraphics*[width=18cm]{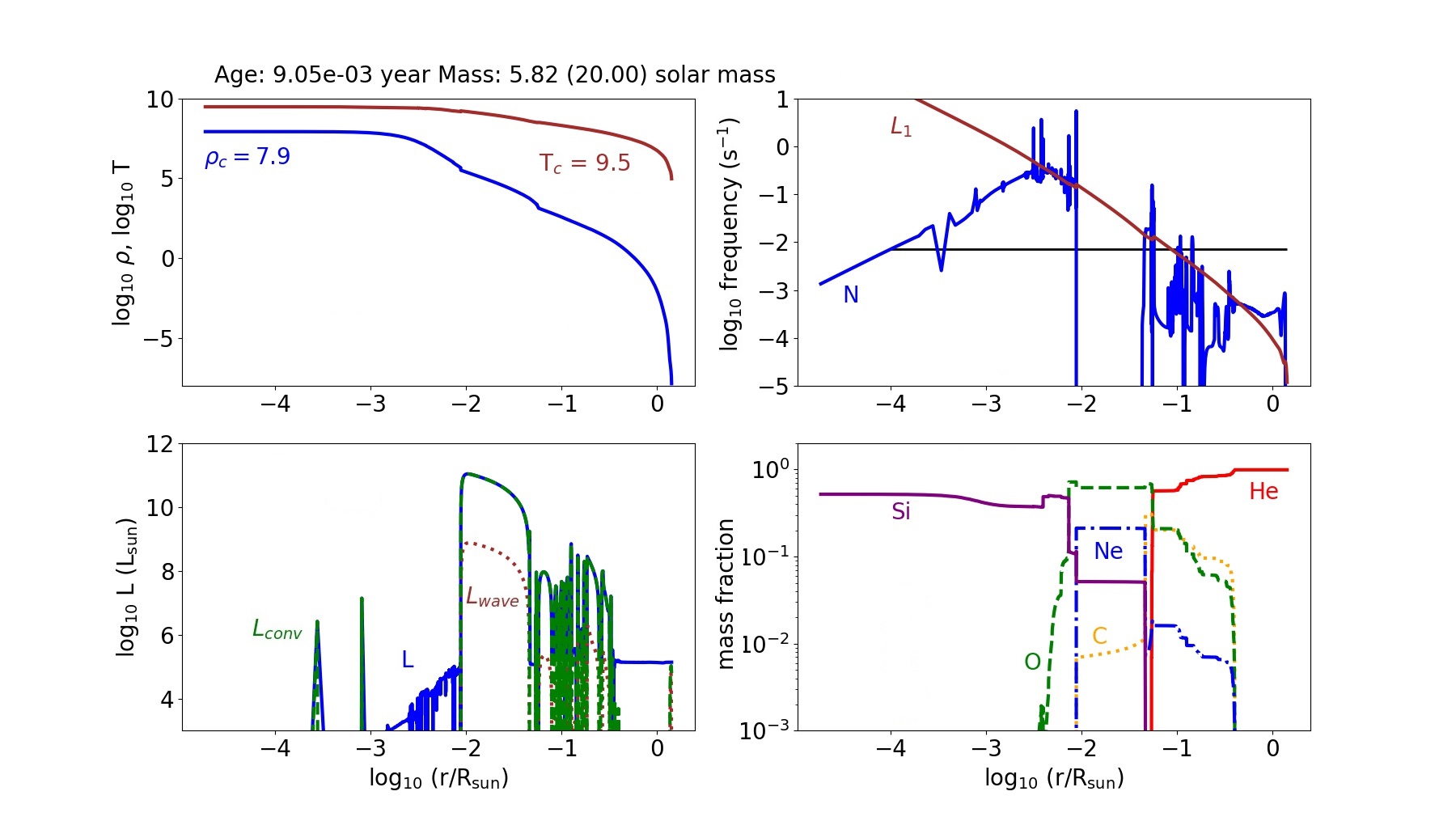}
\caption{The structure of a H-poor model with $M_{\rm ZAMS} = 20~M_{\odot}$ and $Z = 0.007$ taken when the wave heating is strongest at $\sim \! 10^{-2}$ year before collapse. (top left panels) the density and temperature profiles, 
(top right panel) the Lamb frequency (brown line) and Brunt-V{\"a}is{\"a}l{\"a} frequency (blue line), together with the wave frequency (black line), (bottom left panel) the luminosity (blue line), wave luminosity (brown line) and convective luminosity (green line) profiles of the star, and 
(bottom right panel) the chemical abundance profile for He (red), C (orange), O (green), Ne (blue) and Si (purple).}
\label{fig:M20Z0007D05noH_profile}
\end{figure*}

To further diagnose the origin of the different wave energy deposition rates of the models above, we plot in Figure \ref{fig:M20Z0007D05noH_profile} the density and temperature profiles (top left panel), wave propagation diagram (top right panel), the luminosity profile (bottom left panel) and the chemical abundance profile (bottom right panel) of the  $20~M_{\odot}$ model at $\sim 9 \times 10^{-3}$ year from its onset of gravitational collapse. The choice of the model age overlaps with the time where its wave deposition energy is at its maximum. In this model, most of the wave energy arises from the energetic convective shell at $r \sim 10^{-2} \, R_\odot$, which burns carbon, oxygen, and neon and carbon at this snapshot. A recent ingestion of carbon from above has increased the burning luminosity and wave frequency.
The Brunt-V{\"a}is{\"a}l{\"a} frequency and the Lamb frequency profiles indicate the source of the high wave transmission. The outgoing wave frequency is about $10^{-2}$ s$^{-1}$. Waves of this frequency see only a narrow evanescent layer at $r \sim 10^{-1} \, R_\odot$ separating the core from the envelope.
As a result, a large fraction of the waves can escape into the envelope and arrive the surface for heat deposition. 

\subsection{Shell Merger}
\label{sec:caveat}

Another important feature for massive star evolution is the occurrence of convective shell merger events, driven by convective boundary mixing \citep{Collins2018,Davis2019,Andrassy2020, Yadav2020}.
Shell mergers not only change the pre-collapse stellar structure of the massive star, but also provide unconventional thermodynamic conditions for the synthesis of minor elements such as Cl, K and Sc \citep{Ritter2020}.
In our case, convective shell mergers drive vigorous nuclear burning, convection, and wave energy generation, so they can lead to a wave-driven outburst \citep{Wu2020}.

\begin{figure*}
\centering
\includegraphics*[width=8cm]{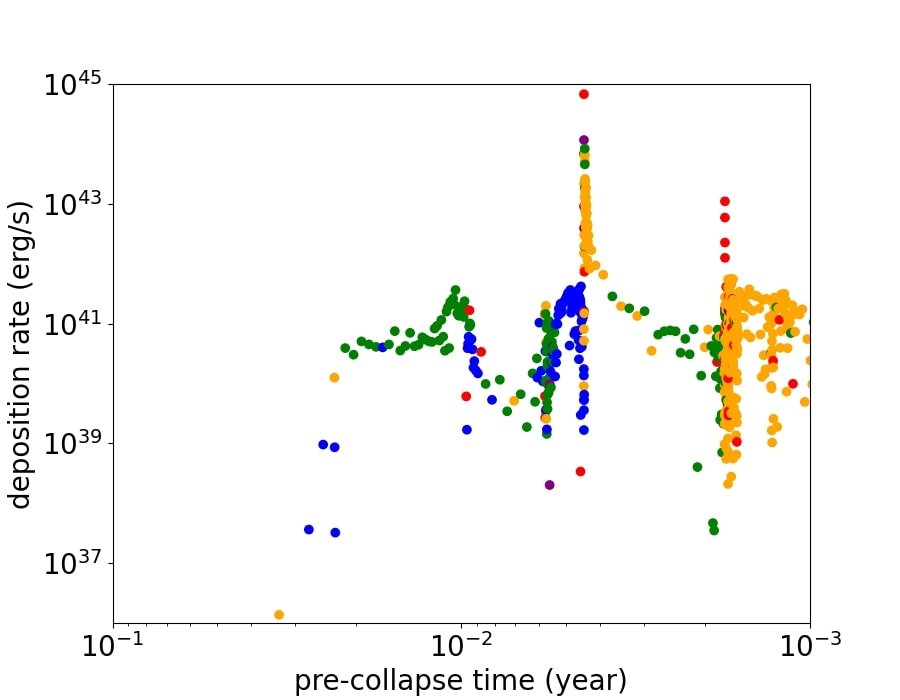}
\includegraphics*[width=8cm]{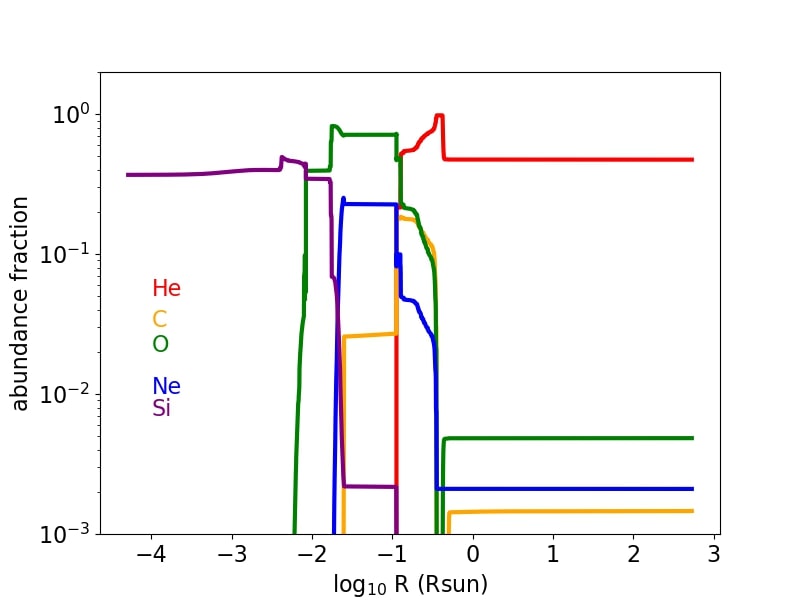}
\includegraphics*[width=8cm]{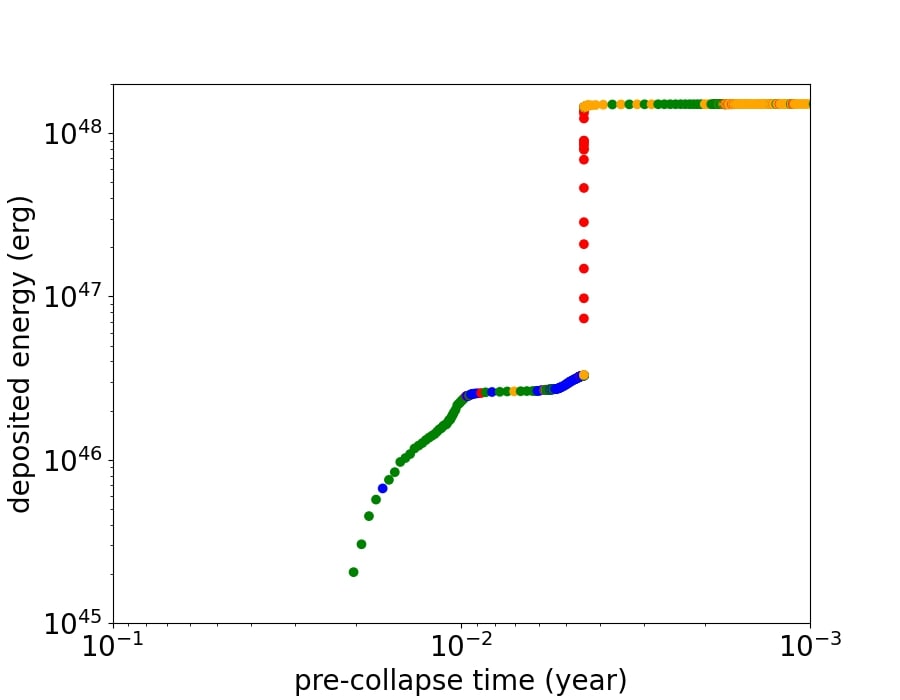}
\includegraphics*[width=8cm]{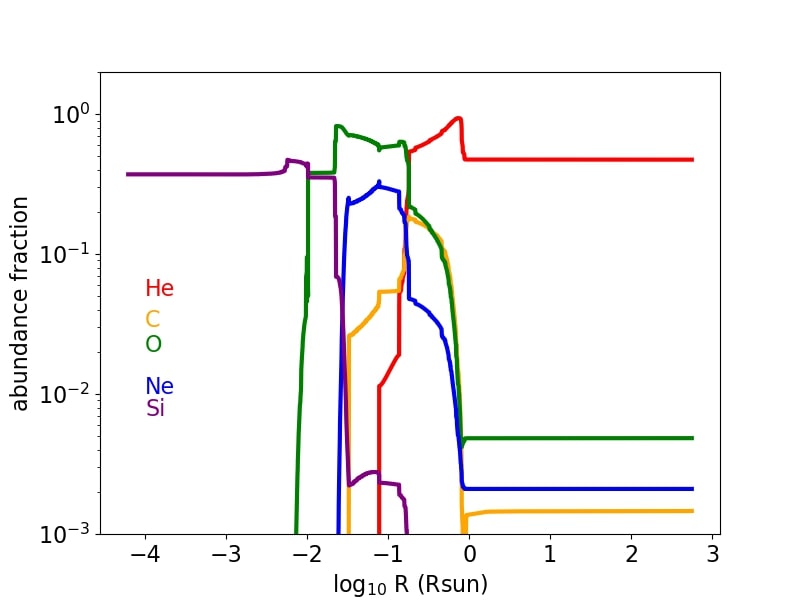}
\caption{
(top left panel) The wave energy deposition rate for the model with $M_{\rm ZAMS} = 45~M_{\odot}$, $Z = 0.02$, with the same colors as Figure \ref{fig:M60Z002D05noH_evol}. (bottom left panel) Same as the top left panel but for the cumulative deposited energy. 
(top right panel) The chemical abundance profile before the merger event of the same model. 
(bottom right panel) Same as the top right panel but after the merger event. }
\label{fig:merger_profile}
\end{figure*}

Our $45 \, M_\odot$ H-rich model exhibits a shell merger that could generate a wave-driven outburst, as shown in Figure \ref{fig:merger_profile}.
When the merger event occurs at 0.005 year before collapse, the energy deposition rate increases by about 3 orders of magnitude. Even though it has a duration less than 0.001 year, more than 90 \% of the deposited energy of $\sim \! 10^{48} \, {\rm erg}$ comes from this event. By examining the chemical abundance profiles before and after the shell merger, we see that He has been dredged down below $10^{-1}~R_{\odot}$ during the merger. This occurs when the convective mixing becomes strong during C-shell burning, causing a merger with the overlying He-burning shell that drags the He-rich material to the actively burning C-shell.
This drives intense nuclear energy via $\alpha$-capture reactions that further power convective motion, and wave excitation. The wave frequency also increases due to the larger convective velocities. Consequently, the wave heating rate of the envelope increases by a few orders of magnitude.

\section{Response of the Envelope}
\label{sec:response}

\subsection{Connection to Rapid Transients}
\label{sec:transient}

Here we study the hydrodynamical response of the envelope due to wave energy deposition. In \cite{Leung2021SN2018gep}, we demonstrated that how fast the energy is deposited affects the envelope expansion. When the energy deposition timescale is shorter than the dynamical timescale, the excited envelope develops a shock and ejects mass in the form of a pulse. On the other hand, when the energy deposition timescale is long, the envelope gradually expands and a steady wind can be driven.

\begin{figure*}
\centering
\includegraphics*[width=18cm, height=10cm]{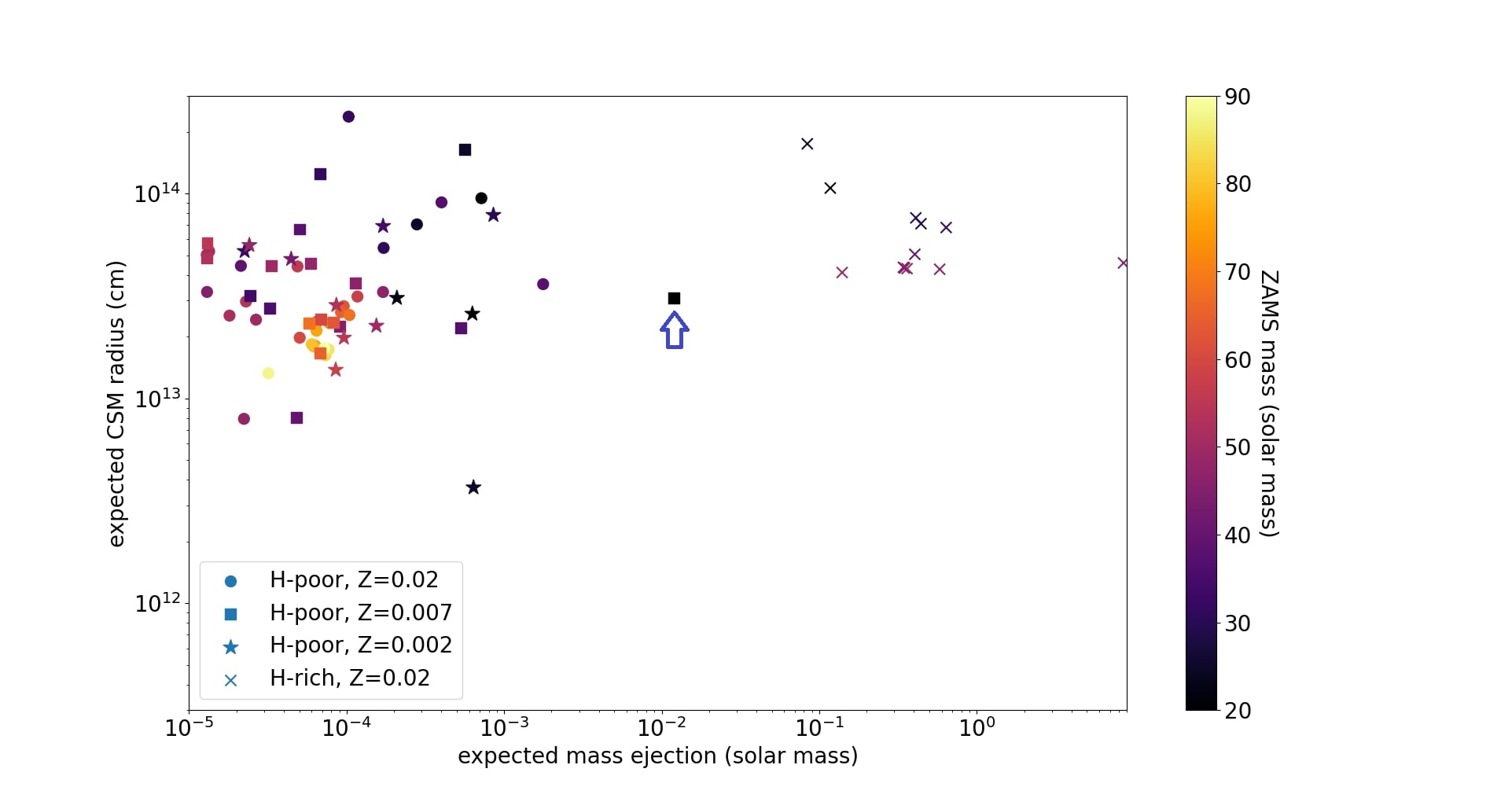}
\caption{
The expected CSM radius against CSM mass based on the energy deposition history for stellar models explored in this work. CSM masses should be considered as upper limits for the H-rich models. The arrow indicates the model for which hydrodynamic and radiative transfer simulations are performed in Figures \ref{fig:wave_hydro}- \ref{fig:rad_tran}.}
\label{fig:expected_CSM}
\end{figure*}

To estimate how much mass can be ejected and its trajectory, we use the total escaped energy $E_{\rm esc}$ and the duration $t_{\rm dep}$. If the wave energy is used efficiently to eject mass, the ejected mass $M_{\rm ej}$ would be approximately
\begin{equation}
\label{mcsm}
    M_{\rm ej} \sim \frac{E_{\rm esc} R}{G M_{\rm pre-SN}},
\end{equation}
where R is the radius of the pre-explosion star. Equation \ref{mcsm} overestimates the ejecta mass in H-rich stars \citep{Fuller2017,Linial2021}, so our estimates for H-rich stars should be taken as upper limits. Similarly, if mass is ejected at the star's escape speed, the CSM radius $R_{\rm CSM}$ would be
\begin{equation}
\label{rcsm}
    R_{\rm CSM} \approx t_{\rm dep} v_{\rm ej} \sim t_{\rm dep} \sqrt{\frac{G M}{R}}.
\end{equation}
\cite{Fuller2018} found that wave heating does efficiently eject mass in H-poor stars, and that the outflow velocity is slightly larger than the escape speed, such that equations \ref{mcsm} and \ref{rcsm} are reasonable estimates. 

In Figure \ref{fig:expected_CSM} we plot $R_{\rm CSM}$ against $M_{\rm ej}$ for all the models computed in this work. The data points cluster into two groups: the H-rich models and the H-poor models. Due to their small binding energy, H-rich models are expected to eject as much as $\sim \! 0.5 M_\odot$, though the true ejected mass may be substantially smaller. If mass is ejected by an outgoing shock wave, \cite{Linial2021} argued that the ejecta mass will likely be either nearly zero or a large fraction of the H-envelope, so more detailed calculations should be performed for reliable estimates of the mass loss in H-rich models.

The H-poor models exhibit a wide range of $M_{\rm ej}$ and $R_{\rm CSM}$ due to their varying wave heating rates and time scales. Most models center around $M_{\rm ej} \sim 10^{-4}~M_{\odot}$ and $R_{\rm CSM} \sim 3 \times 10^{13}$ cm. There are some outliers (corresponding to outliers in Figure \ref{fig:global_Edep}) which have larger ejected mass ($\sim \! 10^{-2}~M_{\odot}$) or a very large $R_{\rm CSM} \sim 10^{14}$ cm. These are models with large wave energies or long wave heating time scales, respectively. The distribution of the data points do not depend strongly on the initial metallicity.  

As shown in \cite{Leung2020COW, Leung2021SN2018gep}, an ejected mass of $\sim \! 10^{-2}$ -- $10^{-1}~M_{\odot}$ with a radius of $10^2$ - $10^{4}~R_{\odot}$ can power luminous SNe via CSM interaction, which explains some rapidly brightening transients with a rise time of $\sim \! 10$ days and a peak luminosity $\sim \! 10^{44}$ erg s$^{-1}$. The expected $M_{\rm CSM}$ is much smaller than the required values for most of our models, so we expect rapidly rising H-poor transients to be uncommon (if the CSM is generated by wave heating).
The exact structure of the CSM and its impact on the explosion light curve require detailed hydrodynamics and radiative transfer calculations, which we perform for a few of our models. 

\subsection{Hydrodynamical Simulations}

Having understood the wave deposition history, we may repeat the procedure outlined in \cite{Leung2021SN2018gep} of depositing the energy to the outer part of the star. In that work, the wave luminosity and duration are chosen from analytic estimates. In this work, we directly use the wave energy deposition history recorded from our stellar models above.
This approach provides a more accurate prescription to study mass ejection and remove extraneous parameters.

Similar to our previous work, we excise the interior, keeping only about $1 ~M_{\odot}$ interior of the C-shell, so that the simulation is not limited by the small time step during the advanced burning in the core.
In our case, the expected mass loss is so small $(< 10^{-2}~M_{\odot})$ that we do not expect the surface motion to feedback on the evolution of the core. 
We add the wave energy as an additional energy source in the stellar models, deposited in the envelope according to Eq. (1) in \cite{Leung2021SN2018gep}. This accounts for acoustic wave energy dissipation via weak shocks and radiative damping. For detailed implementation we refer readers to our previous work.

\begin{figure*}
\centering
\includegraphics*[width=8.5cm]{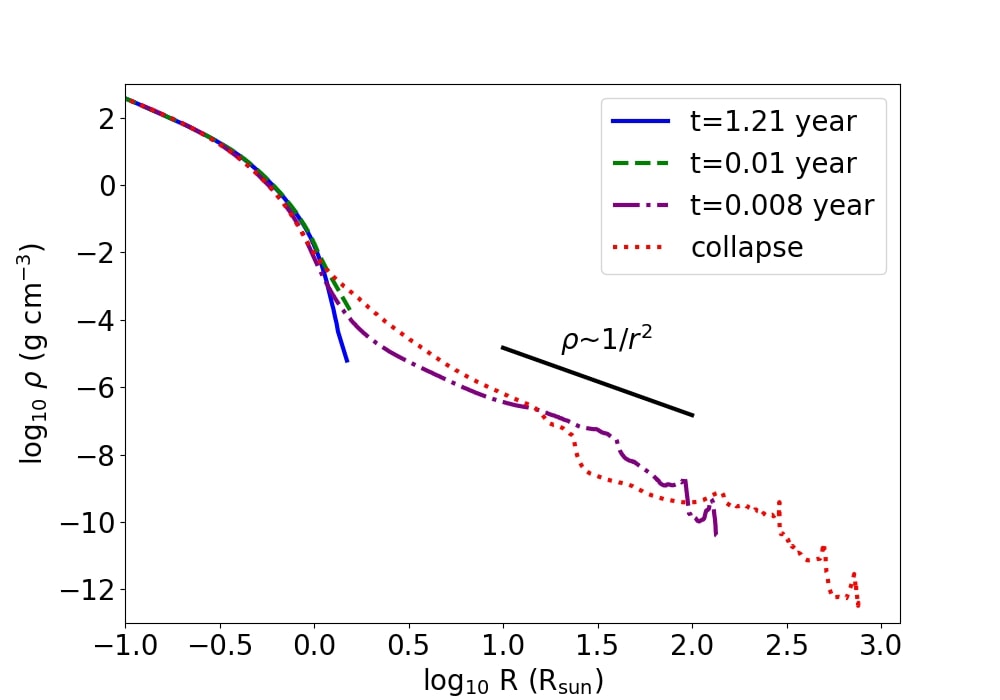}
\includegraphics*[width=8.5cm]{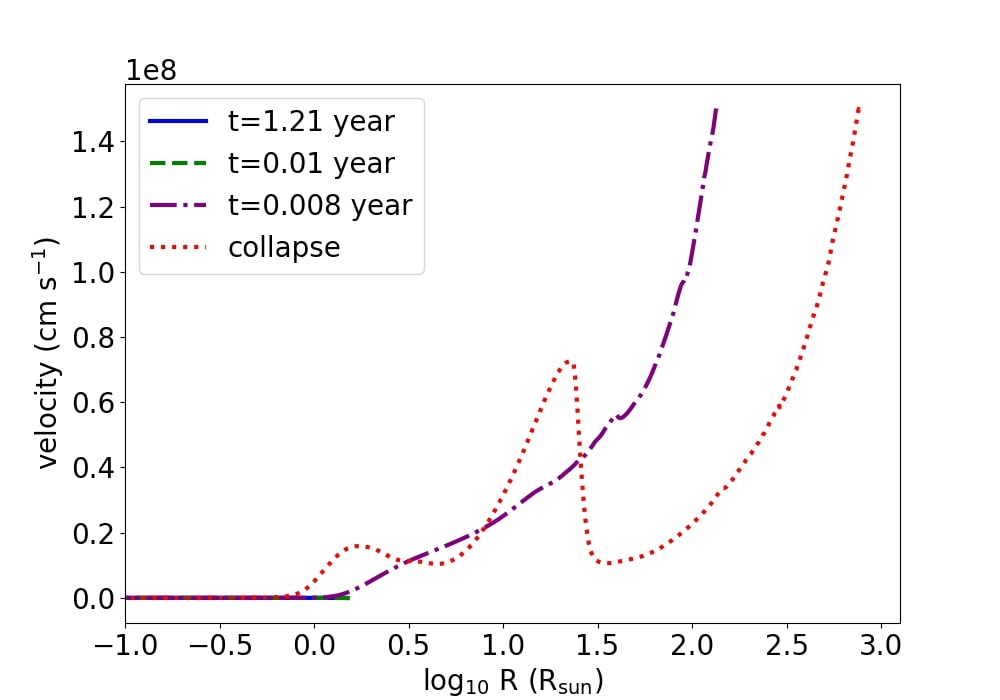}
\includegraphics*[width=8.5cm]{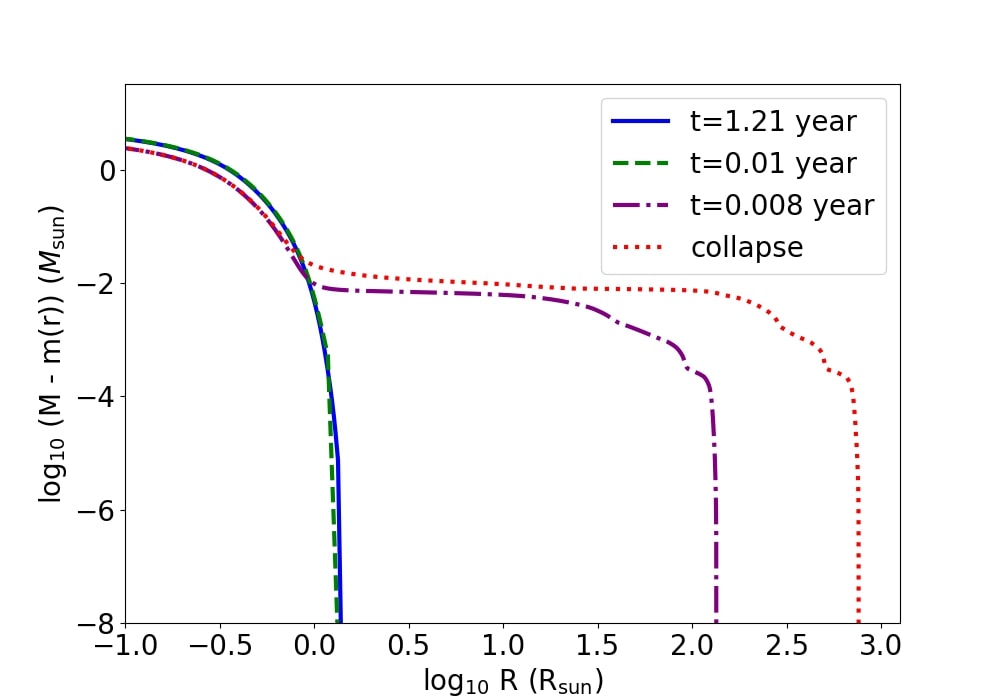}
\includegraphics*[width=8.5cm]{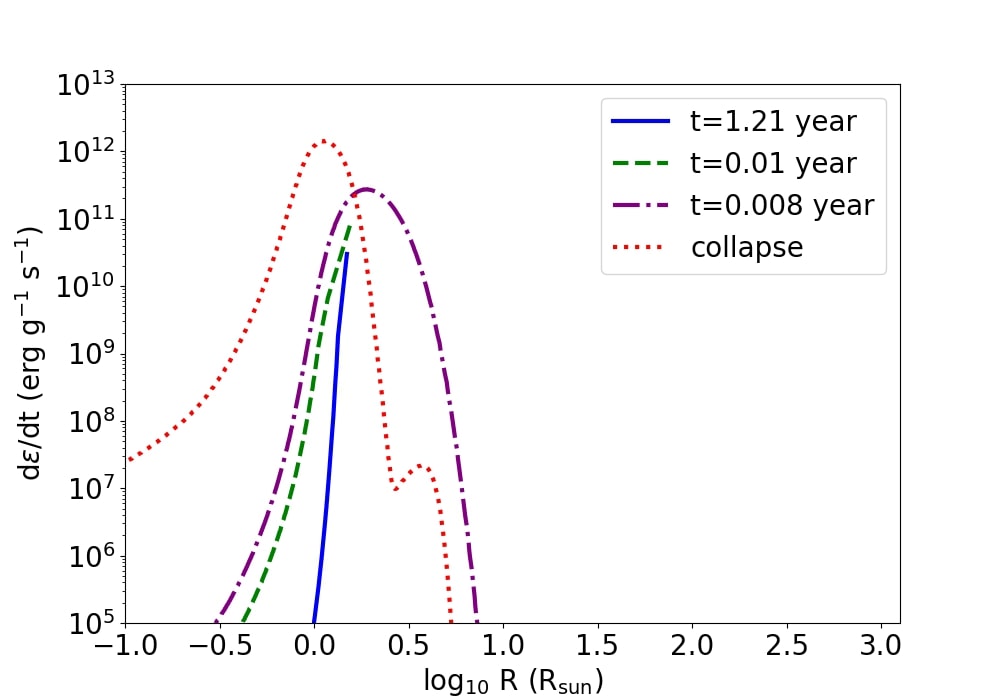}
\caption{
(top left panel) The density profile of the model with $M_{\rm ZAMS} = 20~M_{\odot}$ and $Z = 0.007$ at the beginning of hydrodynamical simulation (blue solid line), 0.01 year before collapse (green dashed line), 0.008 year (purple dashed line) and at the onset of gravitational collapse (red dotted line). (top right panel) Same as the top left panel but for the velocity. (bottom left panel) Same as the top left panel but for the external mass. (bottom right panel) Same as the top left panel but for the heating profile. }
\label{fig:wave_hydro}
\end{figure*}

For demonstration, we consider the H-poor model with the largest mass ejection shown in Section \ref{sec:transient}, with $M_{\rm ZAMS} = 20~M_{\odot}$ and $Z = 0.007$. In Figure \ref{fig:wave_hydro} we plot the density, velocity, mass, and wave heating profiles at different times. The  hydrodynamical simulation begins when the total wave heating energy exceeds $10^3~L_{\odot}$ for the first time. We see that until 0.01 year before collapse, the low energy deposition rate does not lead to significant motion in the envelope. When the Ne-shell burning increases the wave heating rate, we see that the envelope quickly expands, reaching a radius $\sim \! 10^{2.5}~R_{\odot}$. The outermost velocity can reach $\sim  \! 10^8$ cm s$^{-1}$. However, we emphasize that because of the low density, the ejected mass only amounts to $\sim \! 10^{-2}~M_{\odot}$, which is made of mostly He. The surface luminosity can temporarily increase by 2 orders of magnitude to  $L \sim \! 10^{7}~L_{\odot}$.
The density profile of the ejected mass is slightly steeper than a $r^{-2}$-scaling. 

Our analytical mass ejection and timescale estimates match the expected CSM features fairly accurately for the H-poor model with a large $E_{\rm dep}$. For our example, our analytic formulae predict that $M_{\rm CSM}, R_{\rm CSM} \approx \! (0.01~M_{\odot}, 300~R_{\odot})$, whereas the simulation has $M_{\rm CSM}, R_{\rm CSM} \approx \! (0.007~M_{\odot}, 758~R_{\odot})$. However, most of the ejected mass is located below $\sim \! 300 \, R_\odot$ so the analytical estimate is actually quite good. When the expected mass ejection is very small, the analytic formula can overestimate the mass because the energy deposition becomes more concentrated in matter near the stellar surface. 

\begin{figure}
\centering
\includegraphics*[width=8.5cm]{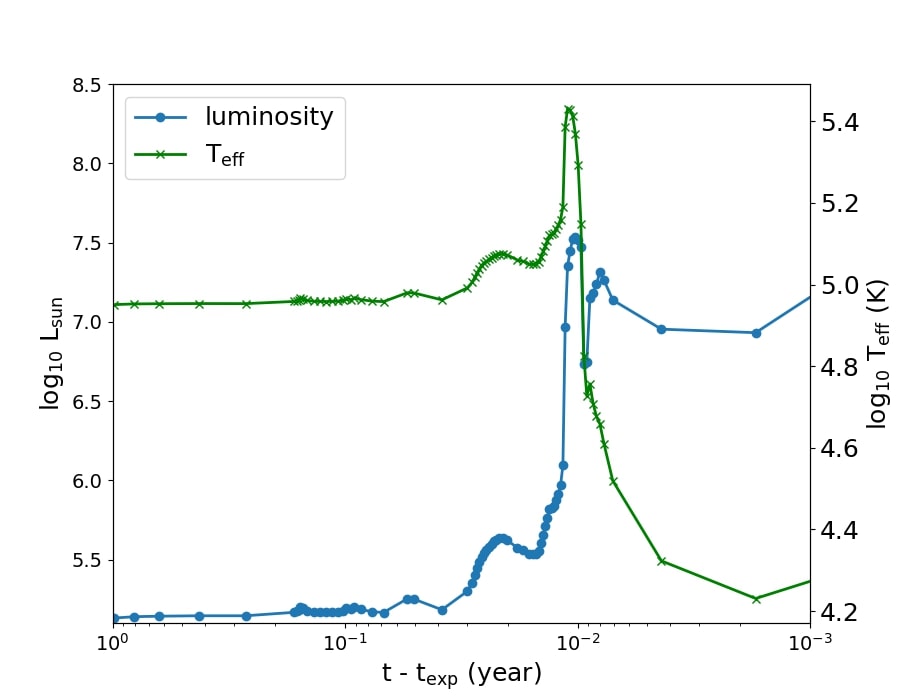}
\caption{
The time evolution of the surface luminosity and effective temperature for the outbursting model shown in Figure \ref{fig:wave_hydro}. }
\label{fig:wave_lumin}
\end{figure}

We plot in Figure \ref{fig:wave_lumin} the time evolution of the surface luminosity for the same model as Figure \ref{fig:wave_hydro}. When the wave heating peaks about 0.01 year before collapse, the surface luminosity increases by a factor of $\sim \! 100$ to reach a peak luminosity of nearly $3 \times 10^7 \, L_\odot$. The temperature initially increases by a factor of a few, but decreases after the initial peak as the photospheric radius moves outward into an extended optically thick wind. The same behavior was seen in \cite{Fuller2018}. High-cadence photometric surveys may be able to detect these progenitor outbursts, like that of SN 2018gep \citep{Ho2020}.

\section{Radiative Transfer of Representative Models}
\label{sec:radtrans}

With the hydrodynamic stellar evolution models, we obtain realistic models of the ejected CSM density profile at the onset of collapse. We use this information to compute the light curve of the subsequent supernova using the one-dimensional radiative transfer hydrodynamics code SNEC \citep{Morozova2015}. The code solves the bolometric radiative transfer in the radiative diffusion limit, with realistic (non-constant) opacity, making it ideal to study the initial shock-cooling part of the light curve, before the ejecta becomes optically thin. 

In the radiative transfer simulations, we choose a few model parameters, including the inner mass cut $M_{\rm cut}$, the explosion energy $E_{\rm exp}$, and the $^{56}$Ni mass  $M_{\rm Ni}$. The simulations are initiated from our stellar progenitor models by excising the Si core (up to $M_{\rm cut}$) and depositing the explosion energy $E_{\rm exp}$ (in units of $10^{51}$ erg) and $^{56}$Ni with a mass $M_{\rm Ni}$ (in units of $M_{\odot}$). We choose $M_{\rm cut}$ to be the Si-core mass of the collapsing model. Explosion energies of $10^{51} - 3 \times 10^{51}$ erg and $^{56}$Ni masses of $0.03-0.3 \, M_\odot$ are used to span the range from typical to energetic Type Ib SNe. 
The CSM mass and radius are taken from the hydrodynamical simulation, but the density profile has been modified to have an $r^{-2}$-dependence to avoid numerical problems that arise due to sharp density changes. 

\begin{figure*}
\centering
\includegraphics*[width=8cm]{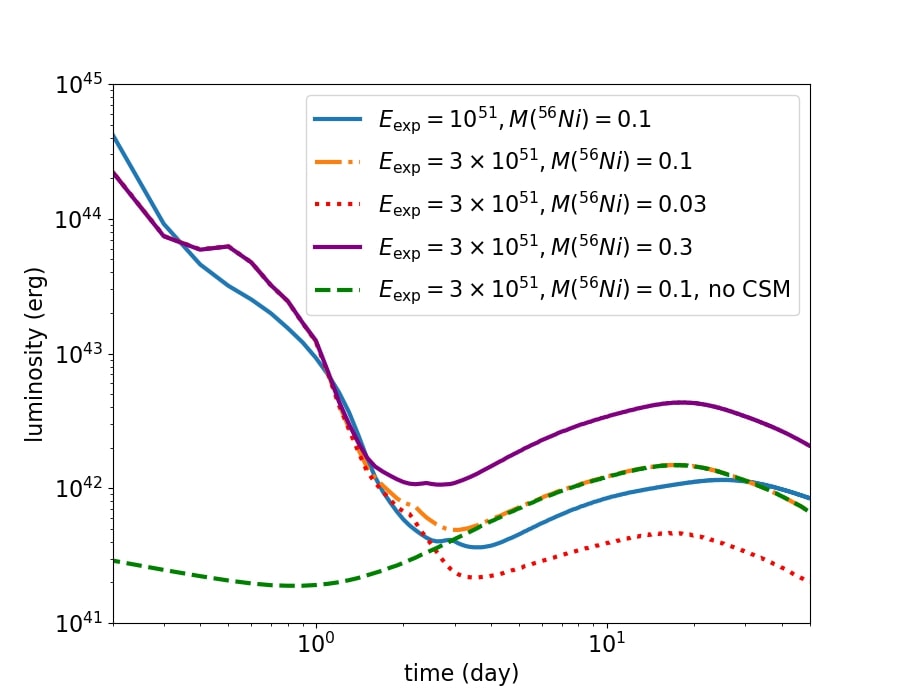}
\includegraphics*[width=8cm]{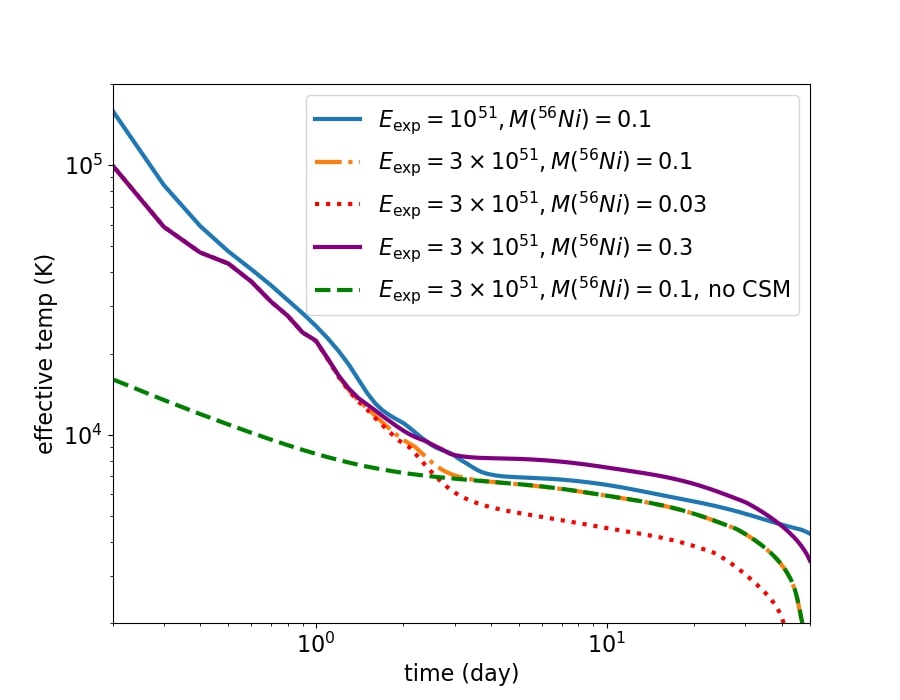}
\includegraphics*[width=8cm]{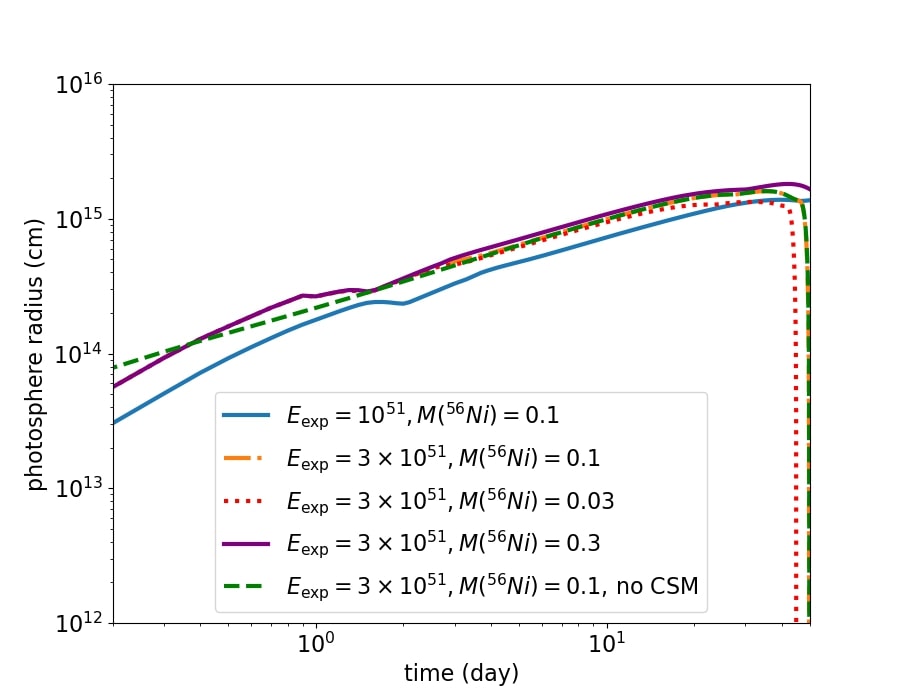}
\includegraphics*[width=8cm]{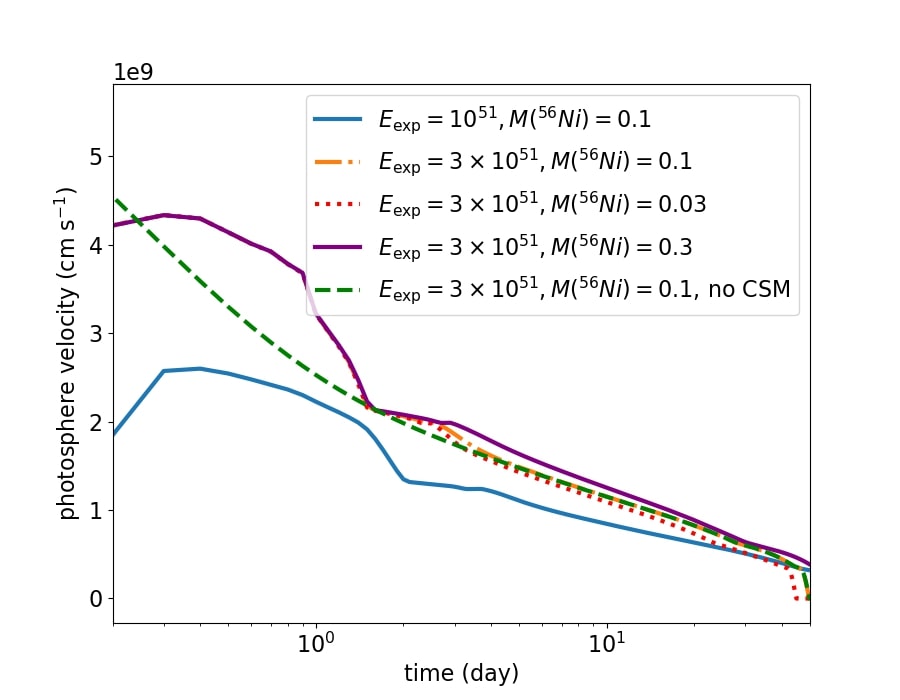}
\caption{
(top left panel) The bolometric light curves of supernova models from the He star model in Figure \ref{fig:wave_hydro} as a function of time after explosion.
Models are named by the format $E_{\rm exp}$-$M(^{56}$Ni) as described in the text.
(top right panel) Same as the top left panel but for the effective temperature. 
(bottom left panel) Same as the top left panel but for the photosphere radius profile. 
(bottom right panel) Same as the top left panel but for the photosphere velocity. }
\label{fig:rad_tran}
\end{figure*}

In Figure \ref{fig:rad_tran} we plot the radiative transfer results of our representative wave-driven mass loss model from the preceding section.
A control model with no CSM is added to highlight the effect of the CSM on the supernova appearance.
The light curve has a very sharp rise and fall due to shock cooling emission from the CSM, similar to the light curves of many type IIb SNe. In type IIb SNe, the shock cooling emission comes from the low-mass extended H envelope ($M \sim 0.01 \, M_\odot$, $R \sim 200 \, R_\odot$). Our models have CSM with a similar mass and radius (though it is composed of He instead of H) and the early light curves thus look quite similar. Because of the low $M_{\rm CSM}$, the shock cooling time is short ($\sim$ 3 days), but the bolometric luminosity can be as high as $\sim 10^{44}$ erg. The high temperature ($T > 3 \times 10^4 \, {\rm K}$) at the peak implies the shock cooling would be associated with bright UV emission.

The effective temperature and photosphere velocity also show a rapid rise and fall during the early CSM shock cooling phase. After the initial peak, the photosphere recedes below the CSM, and the light curve then resembles a typical Type IIb or Ib SN. Higher $^{56}$Ni will lead to a more significant rise in the luminosity after the shock front cools down. The explosion energy primarily changes how fast the light curve falls after day $\sim 20$.  A lower $^{56}$Ni leads to a faster drop in the effective temperature after shock breakout, while a lower explosion energy makes the effective temperature drop slower.

The photosphere radius increases rapidly at early time as the surface traces the outer layers of the expanding ejecta during the shock cooling phase. After that, its expansion is balanced by the recession in the Lagrangian sense such that the photospheric radius pleateaus at $\sim \! 10^{15} \, {\rm cm}$. The photospheric velocity converges to $\sim \! 10^4$ km s$^{-1}$ until the ejecta become transparent, with higher explosion energies creating higher photospheric velocities as expected. The ejecta becomes transparent to optical radiation at about 45--60 days after explosion, at which point SNEC's effective temperature, photospheric radius, and velocity become unreliable.


\section{Discussion}
\label{sec:discussion}

\subsection{Rapidly rising and fading transients}

Our wave heating models can be compared to rapidly rising and fading transients \citep[e.g.,][]{drout:14}
which may be powered by CSM interaction. For the large majority of our models, the wave heating is insufficient to eject enough mass to power observed rapidly evolving transients. 
The typical mass scale of wave-driven outbursts of $\sim \! 10^{-4} \, M_\odot$ can only sustain shock cooling emission for $\sim \! 10^{-1} \, {\rm d}$, far shorter than the typical evolution times of $\sim \! 10 \, {\rm d}$. The upper end of the distribution explored in this work ejects $\sim \! 10^{-2} \, M_\odot$ which may sustain shock cooling for $\sim \! 1 \, {\rm d}$, still too short compared to the current population of observed rapidly evolving transients. However, our models predict that a significant fraction of ``ordinary" type Ib/c SNe will exhibit a bright but very brief phase of CSM interaction within hours of explosion.

Our models also shed light on the rapid rise and fall of the light curve in heavily stripped type Ib/c SNe such as iPTF14gqr \citep{De2018}, iPTF16hgs \citep{de:18b}, and SN 2019dge \cite{Yao2020}. The short-lived shock-cooling for iPTF14gqr indicated a CSM mass and radius of $M_{\rm CSM} \sim 0.01 \, M_\odot$ and $R_{\rm CSM} \sim 3 \times 10^{13} \, {\rm cm}$, similar to that of our most energetic H-poor model in Figure \ref{fig:expected_CSM}. Hence, wave driven mass loss could possibly account for the extended envelope of that event. For the other two, the authors inferred
$M_{\rm CSM} \sim \! 0.1 \, M_\odot$ and $R_{\rm CSM} \sim 10^{12} \, {\rm cm}$ for iPTF16hgs, and $M_{\rm CSM} \sim \! 0.1 \, M_\odot$ and $R_{\rm CSM} \sim \! 3 \times 10^{12} \, {\rm cm}$ for SN 2019dge. Those masses are likely too large to be created by wave driven mass loss, and the radii are likely too small. Hence, an inflated He envelope as is naturally expected in low-mass He star SN progenitors may be a more likely explanation for those events. All of these events were heavily stripped SNe (likely in binary systems) arising from low-mass progenitors whose structures are quite different from the higher-mass He stars studied here. Future calculations should investigate H-poor low-mass He stars, whose wave heating rates may be increased by degenerate neon/oxygen/silicon burning \citep{Wu2020}.

\subsection{Shock breakout emission}

Even though the small amount of CSM predicted by our models will not have a large impact on the optical light curves of typical type Ib/c SNe, it may affect the shock breakout emission. Even a small amount of optically thick mass ($M \gtrsim 10^{-6} \, M_\odot$) above the photosphere may be sufficient to affect the shock breakout duration and its X-ray spectrum.

Our models are roughly consistent with the enhanced CSM density inferred around the progenitor of type Ib SN 2008D based on its shock breakout emission \cite{soderberg:08}. \cite{svirski:14} (see also \citealt{balberg:11,ioka:19,ito:20}) demonstrated that the CSM density at $R \lesssim 10^{14} \, {\rm cm}$ must have been $\sim \! 20$ times larger than expected for Wolf-Rayet stars. This entails $\dot{M} \sim 10^{-3} \, M_\odot/{\rm yr}$ for a wind mass loss rate of $5 \times 10^{-5} \, M_\odot/{\rm yr}$. Our typical models have $R_{\rm CSM} \lesssim 10^{14} \, {\rm cm}$, consistent with the shock breakout constraint. They also have $M_{\rm CSM} \sim \! 10^{-4} \, M_\odot$, ejected on time scales of $t_{\rm CSM} \sim 10^{-2} \, {\rm yr}$, corresponding to mass loss rates of $\sim 10^{-2} \, M_\odot/{\rm yr}$. This is only slightly larger than the inferred late-time mass loss rate of SN 2008D. More detailed modeling of shock-breakout emission (which is not accurately captured by our SNEC models) can determine whether wave-driven mass can account for the shock breakout signal of 2008D. In any case, we predict that longer-than-expected shock breakout emission may be very common in H-poor SNe, which may be tested by future wide-field UV surveys \citep{sagiv:14}.

\subsection{Comparison with Literature Work}
\label{sec:comparison}

Our work can be compared with other studies of outbursts driven by sudden energy deposition. In \cite{Kuriyama2020}, energy is deposited in the stellar envelope with a given deposition rate and duration, and the amount of energy is scaled with the binding energy of the envelope. The resultant mass loss ranges from $\sim 10^{-3} - 1~M_{\odot}$.
For red supergiants losing mass via a shock, the actual mass loss sharply drops when the ratio of deposited energy to binding energy is less than roughly 0.5, as explained in \cite{Linial2021}. 
For compact stars with mass loss via super-Eddington winds, most of the deposited energy is used to lift mass out of the gravitational potential, so the total mass lost is closer to our estimate from equation \ref{mcsm}, as explained in \cite{Quataert2016}.

In \cite{Ouchi2021}, a moderate heating rate of of $\sim \! 10^{39}$ erg s$^{-1}$ is added to a $12~M_{\odot}$ star, but with a long-lasting duration of 3 years. The total energy is about $10^{47}$ erg which is close to our H-rich models, but our models suggest that the expected energy is lower for H-poor stars. The deposition time scale of three years is much longer than our models, allowing the star to expand smoothly rather than eject mass via outbursts.
In \cite{Owocki2019} the parametrized energy deposition in a stellar envelope is studied. Half of the envelope energy ($\sim 10^{50}$ erg) is deposited in a blue supergiant model and a mass outburst of $\sim 7~M_{\odot}$ is observed, resembling the mass eruption in the $\eta$-Carinae. Such a high energy deposition is not seen in our series of models, suggesting that other mechanisms are necessary to explain massive outbursts like that of $\eta$-Carinae.

\subsection{Caveats}

In this work we have assumed that the star loses all of its H-envelope due to interactions from its binary companion. This is a good approximation for high-mass or high-metallicity stars where winds will likely remove any residual H-envelope during core He-burning. For lower mass and lower metallicity stars, 
winds might not robustly remove all the H \citep{Goetberg2017,Laplace2020},
and an H-envelope may remain till the end of stellar evolution. The less compact H-envelope would have a lower binding energy, possibly triggering more significant expansion and/or mass loss compared to the pure He case. This scenario should be investigated in future work.

We have used bolometric radiative transfer for computing the light curves, assuming blackbody radiation. However, given the very low CSM mass, the medium becomes transparent at a very early time ($\sim$10 days), which could mean that photon transport depends on a frequency-dependent opacity. In order to robustly model optically-thin ejecta and the associated effective temperature, multi-band radiative transport becomes necessary.

\section{Conclusion}
\label{sec:conclusion}

We have explored the physics and consequences of wave energy transport in an extensive survey of H-poor stripped-envelope massive stars. Our models account for multiple convective nuclear burning shells that generate gravity waves, and how these waves tunnel through the star and deposit energy in the envelope. These models improve upon prior efforts by employing a more realistic spectrum of wavenumbers and by including non-linear damping of waves within the gravity mode cavity in the core of the star. We have surveyed stars with a ZAMS mass from 20 -- 90 $M_{\odot}$ and a metallicity from 0.002 to 0.02. 

We find that wave heating rates in the H-poor stars are somewhat different from H-rich stars. In H-poor stars, in general a smaller amount of the wave energy is able to escape into the surface layers, with typical wave energy deposition of $\sim \! 10^{46}$ erg during the last $\sim \! 10^{-2}$ years of the star's evolution. The smaller escape fraction is caused by a differing stellar structure at the outer edge of the helium core that creates thicker evanescent layers, trapping the waves in the core and increasing their damping via non-linear dissipation.
The majority of the wave energy that does escape arises from convective wave generation by O-burning shells. We find no convective shell mergers in our H-poor models, so energetic outbursts associated with these events did not appear in our suite of models.
The overall wave heating does not vary strongly with the initial metallicity.

Using hydrodynamic stellar models including wave heating, we have also estimated the associated CSM formation structure via wave-driven mass loss. The small wave heating rates of our models in general generates a small CSM mass $\sim 10^{-5}$ -- $10^{-3}~M_{\odot}$, with a wide range of radial extents between $10^{12} - 10^{14}$ cm.
Due to the high surface binding energy of compact He stars, the ejecta mass is much smaller than what is attainable from a H-rich star (which could in principle eject $\sim \! 10^{-1}~M_{\odot}$ prior to its collapse). Hence, in most cases, we do not expect wave-driven outbursts to eject enough mass to cause interaction-powered H-poor SNe, explaining why CSM interaction is not frequently observed in Type Ib/c SNe. Our most energetic models eject $\sim \! 10^{-2} \, M_\odot$ which can power a bright but brief phase of shock cooling emission, which we model using radiative transfer models with SNEC. The rapid rise and fall of the light curve is similar to that observed in some heavily stripped SNe, so this possibility should be explored in future work. We do predict a small amount of CSM around most type Ib/c SNe, which could greatly affect the SN shock breakout signal in UV/X-ray bands.

In future work, our models will be extended to lower mass He stars, where degenerate ignition of Ne, O, and Si may power more energetic wave-driven outbursts \citep{Wu2020} with larger CSM masses and radii. Those types of outbursts remain viable candidates for interaction-powered H-poor SNe such as type Ibn supernovae.

\section{Acknowledgments}

S.C.L thanks the MESA development community for making the code open-sourced and V. Morozova and her collaborators in providing the SNEC code open source. S.C.L. and JF acknowledges support by NASA grants HST-AR-15021.001-A and 80NSSC18K1017. 

\software{MESA \citep{Paxton2011,Paxton2013,Paxton2015,Paxton2017,Paxton2019} version 8118; SNEC \citep{Bersten2011, Bersten2013, Morozova2015} version 1.01; Python libraries: Matplotlib \citep{Matplotlib}, Pandas \citep{Pandas}, Numpy \citep{Numpy}.}

\bibliographystyle{aasjournal}
\pagestyle{plain}
\bibliography{biblio}

\end{document}